\documentclass[a4paper,11pt]{article}
\pdfoutput=1

\usepackage{jheppub}

\usepackage[abs]{overpic}

\usepackage{warpcol}

\usepackage{lineno}
\usepackage{longtable}
\usepackage{lscape}
\usepackage{graphicx}% Include figure files
\usepackage{threeparttable} % 专业表格宏包
\usepackage{booktabs}
\usepackage{dcolumn}% Align table columns on decimal point
\usepackage{bm}% bold math
\usepackage{rotating}
\usepackage{epstopdf}
\usepackage{color}
\usepackage{verbatim} %for comment
\usepackage{multirow}
\usepackage[abs]{overpic}
\usepackage{amsmath}
\usepackage{mathrsfs}
\usepackage{amssymb}
\usepackage{subfigure}
\usepackage{xspace}
\usepackage{float}
\usepackage{hyperref}
\usepackage{booktabs}
\usepackage{graphicx}
\usepackage{lipsum}
\usepackage{caption}
\usepackage{afterpage}
\hypersetup{colorlinks=true, linkcolor=blue, anchorcolor=blue, citecolor=blue}

\newcommand{\PreserveBackslash}[1]{\let\temp=\\#1\let\\=\temp}
\newcolumntype{C}[1]{>{\PreserveBackslash\centering}p{#1}}
\newcolumntype{R}[1]{>{\PreserveBackslash\raggedleft}p{#1}}
\newcolumntype{L}[1]{>{\PreserveBackslash\raggedright}p{#1}}

\newcommand{\EE}{e^+e^-}

\newcommand{\BESIIIorcid}[1]{\href{https://orcid.org/#1}{\hspace*{0.1em}\raisebox{-0.45ex}{\includegraphics[width=1em]{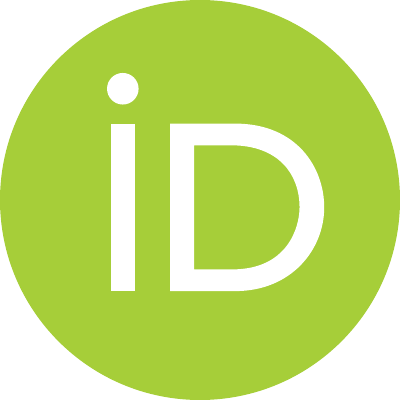}}}}

\usepackage[T1]{fontenc}

\RequirePackage{lineno}
\usepackage{epstopdf}

\title{\boldmath First measurement of the cross sections for $e^{+}e^{-}\to K^{0}K^{-}\pi^{+}J/\psi+c.c.$ at $\sqrt{s}$ from 4.396 to 4.951 GeV}

\collaborationImg{\includegraphics[width=.12\textwidth,origin=c,angle=90]{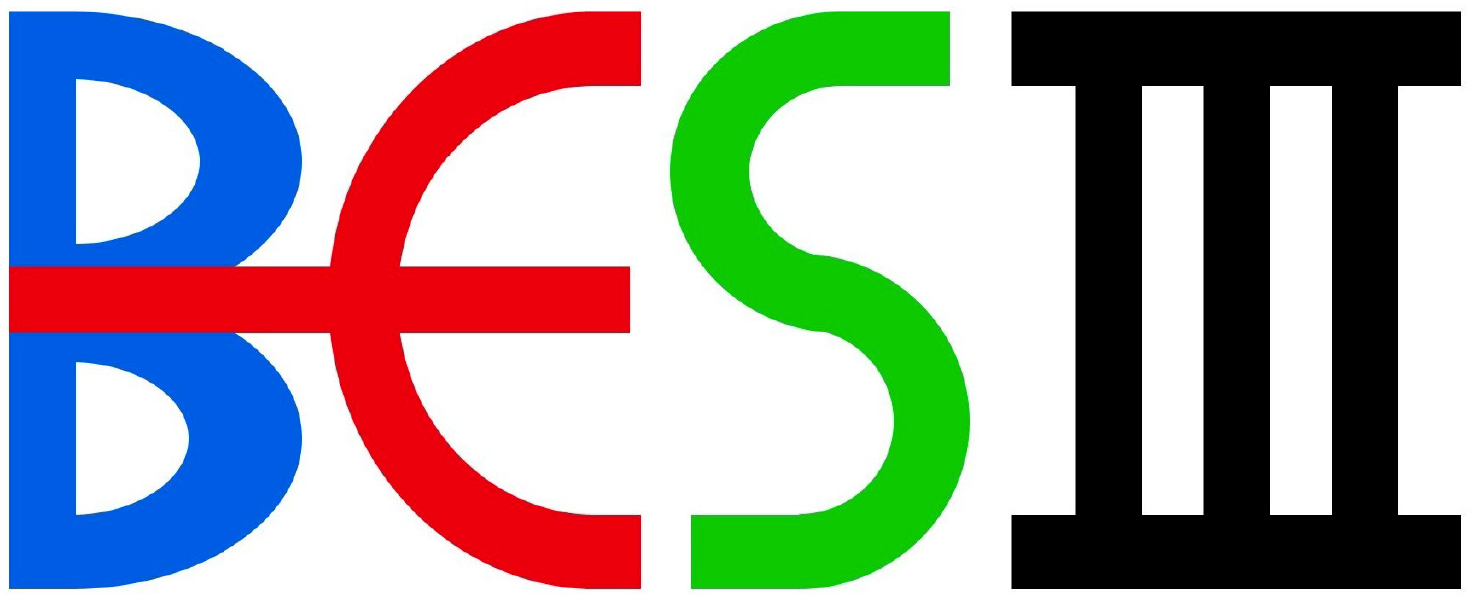}}

\collaboration{BESIII Collaboration}


\emailAdd{besiii-publications@ihep.ac.cn}

\arxivnumber{1234.5678}

\abstract{Using $e^+e^-$ collision data at 19 center-of-mass energies ranging from $4.396$ to $4.951~\mathrm{GeV}$ corresponding to a total integrated luminosity of $8.86~{\rm fb}^{-1}$ collected by the BESIII detector, the process $e^+e^-\to K^{0}K^-\pi^+ J/\psi+c.c.$ is observed for the first time, with a statistical significance of $9.4\sigma$ summing up all the data samples. For this process, the cross section and the upper limit at the $90\%$ confidence level are reported at each of the 19 center-of-mass energies.~No statistically significant vector structures are observed in the cross section line shape, nor are any intermediate states of $K\pi$, $K\bar{K}$, $K\bar{K}\pi$, $KJ/\psi$, $\pi J/\psi$, and $K\pi J/\psi$ seen at individual energy points or in the combined data sample.}

\begin{document}
\maketitle
\flushbottom

%\linenumbers
%%%%%%%%%%%%%%%%%%%%%%%%%%%%%%%%%%%%%%%%%%%%%%%%%%%%%%%%%%%%%%%%%%%%%%%%%%%%%%
%%%%%%%%%%%%%%%%%%%%%%%%%%%%%%%%%%%%%%%%%%%%%%%%%%%%%%%%%%%%%%%%%%%%%%%%%%%%%%
\section{Introduction}
The charmonium-like states are promising candidates for exotic hadrons, such as tetraquarks, hadronic molecules, and hybrid mesons ~\cite{Yreview,review1,review2,review3,review4,review5}, since they show properties that are not compatible with the conventional $c\bar{c}$ meson spectrum.~These states provide valuable insight into non-perturbative quantum chromodynamics~(QCD).~As the first observed $Y$ state,
the $Y(4260)$ (aka $\psi(4230)$~\cite{PDG2018}) was discovered by the BaBar experiment using the initial state radiation (ISR) process $e^{+}e^{-}\to \gamma_{\rm ISR} \pi^{+}\pi^{-}J/\psi$~\cite{Y4260}. Based on a series of cross section measurements for $e^{+}e^{-}\to$~hadrons, including both hidden-charm and open-charm processes~\cite{intro-Belle-pipijpsi,BESIII:pipijpsi,BESIII:pipijpsi new,intro-BESIII-pipihc2,psi_3686_1,psi_3686_2,psi_3686_3,BESIII:omegaChic0,BESIII:piDDstar1,BESIII:piDDstar2,BESIII:etaJpsi,etajpsi,DsDsja,DsDsjb,DsDsJ,dssdss,kkjpsi,ksksjpsi,depth}, several additional $Y$ states have been observed.~In addition, the excited charmonium states above $4.6~{\rm GeV}/c^2$ predicted by the potential model, such as the $5S$ and $4D$ states, are not yet well established~\cite{p1,p2,p3,p4,p5}. Experimentally, the $Y(4660)$ (aka $\psi(4660)$~\cite{PDG2018}) state was observed by the Belle experiment in the process $e^{+}e^{-}\to \gamma_{\rm ISR} \pi^{+}\pi^{-}\psi(3686)$~\cite{psi_3686_1}, confirmed by the BaBar~\cite{psi_3686_2} and BESIII~\cite{psi_3686_3} experiments.~In addition, enhancements around $4.6{\rm~GeV}$ are observed in the cross section line shapes for both the process $e^{+}e^{-}\to D^{+}_{s}D_{s1}(2536)^{-}+c.c.$ and $e^{+}e^{-}\to D_{s}^{+}D_{s2}^{*}(2573)^{-}+c.c.$~\cite{DsDsja,DsDsjb,DsDsJ},~where $c.c.$ stands for the charge conjugated mode.~The $Y(4500)$, $Y(4710)$, and $Y(4790)$ were discovered by the BESIII experiment in the cross section line shape of the processes $e^{+}e^{-}\to K\bar{K}J/\psi$~\cite{kkjpsi,ksksjpsi,depth} and $e^{+}e^{-}\to D_{s}^{*+}D_{s}^{*-}$~\cite{dssdss}, respectively.~These observations imply some of $Y$ states may harbor strange and charm quark-antiquark ($s\bar{s}$ and $c\bar{c}$) components within their wavefunction configurations.~This motivates the investigation for the decay $Y \to K^{0}K^-\pi^+ J/\psi+c.c.$ as a natural extension of the current research. 

Recently, the BESIII experiment reported the observation of a structure, denoted as $Z_{cs}({3985})^{-}$, with a significance of $5.3\sigma$~\cite{zcs3985a}, decaying into the ($D_{s}^{-}D^{*0} + D_{s}^{*-}D^{0}$) final state.~Evidence for its neutral counterpart, $Z_{cs}({3985})^{0}$~\cite{zcs3985b}, has also been found in the ($D_{s}^{+}D^{*-} + D_{s}^{*+}D^{-}$) system, with a significance of $4.6\sigma$.~In addition, two charged $Z_{cs}~(Z_{cs}(4000)^{+}/Z_{cs}(4220)^{+} \to K^{+}J/\psi)$ candidates were discovered in an amplitude analysis of the $B^{+}\to J/\psi \phi K^{+}$ decay at LHCb~\cite{lhcb}. 
Discussions are ongoing about whether $Z_{cs}(3985)^{+}$ and $Z_{cs}(4000)^{+}$ represent the same state~\cite{kk1,kk2,kk3,kk4,kk5,kk6,kk7,kk8,kk9}, as their masses are comparable, but their decay widths differ by nearly an order of magnitude.~The $Z_{cs}$ states have also been explored in the~$KJ/\psi$~and~$K \psi(3686)$ systems through the processes $e^{+}e^{-}\to K\bar{K}J/\psi$~\cite{depth} and $e^{+}e^{-}\to K\bar{K}\psi(3686)$~\cite{kkpsi1,kkpsi2} by the BESIII collaboration, but no significant signals were observed.~Investigations of those $Z_{cs}$ states through hidden-charm and open-strange channels play a crucial role in establishing their structural nature as molecular states, tetraquarks, or other exotic configurations~\cite{kk4,kk9,zcs1,zcs2,zcs3,zcs4,zcs5}.~The $K^{0}K^-\pi^+ J/\psi+c.c.$ final state also allows searching for $Z_{cs}$ states with quantum numbers different from those of the observed states (such as $J^{P}=0^{+}$). 

In addition, the authors of Refs.~\cite{K4307a,K4307b} predicted an excited $K$ state with hidden charm by solving the Faddeev equations~\cite{K4307c} , denoted as $K^{*}(4307)$.~Its mass and width are $M = 4307~{\rm MeV}/c^{\rm 2}$ and $\Gamma = 18~{\rm MeV}$, and its quantum numbers are $I(J^{P})~=~1/2(1^{-})$.~The $K^{*}(4307)$ is expected to decay predominantly to $K^{*}J/\psi$, and can be searched for via the process $\EE\to K^{0}K^-\pi^+ J/\psi+c.c.$

In this paper, we report the first observation of the process $\EE\to K^{0}K^-\pi^+ J/\psi+c.c.$ using data samples collected at 19 center-of-mass (c.m.) energies ($\sqrt{s}$) from 4.396~to~4.951~GeV, corresponding to a total integrated luminosity of $8.86~{\rm fb}^{-1}$. The c.m. energy of the data sample is measured using the di-muon process for $\sqrt{s}~<~4.6~{\rm GeV}$~\cite{luma,lumc} and the
$\Lambda_c\bar{\Lambda}_c$ pairs in $e^+e^-$ annihilation for $\sqrt{s}~>~4.6~{\rm GeV}$ ~\cite{BESIII:lum}.
The integrated luminosities of these data samples are measured using large-angle Bhabha scattering sample~\cite{lumb,BESIII:lum}. 

\section{BESIII detector and Monte Carlo simulation}

\indent
The BESIII detector~\cite{bes} records symmetric $e^+e^-$ collisions 
provided by the BEPCII storage ring~\cite{Yu:IPAC2016-TUYA01}
in the c.m.~energy range from 1.84 to 4.95~GeV, with a peak luminosity of $1.1 \times 10^{33}\;\text{cm}^{-2}\text{s}^{-1}$ achieved at $\sqrt{s} = 3.773\;\text{GeV}$. BESIII has collected large data samples in this energy region~\cite{Ablikim:2019hff,EcmsMea,EventFilter}. The cylindrical core of the BESIII detector covers 93\% of the full solid angle and consists of a helium-based
 multilayer drift chamber~(MDC), a time-of-flight
system~(TOF), and a CsI~(Tl) electromagnetic calorimeter~(EMC),
which are all enclosed in a superconducting solenoidal magnet
providing a 1.0~T magnetic field.~The solenoid is supported by an
octagonal flux-return yoke with resistive plate counter muon
identification modules interleaved with steel. 
%The acceptance of charged particles and photons is 93\% over $4\pi$ solid angle. 
The charged-particle momentum resolution at $1~{\rm GeV}/c$ is
$0.5\%$, and the 
${\rm d}E/{\rm d}x$
resolution is $6\%$ for electrons
from Bhabha scattering. The EMC measures photon energies with a
resolution of $2.5\%$ ($5\%$) at $1$~GeV in the barrel (end cap)
region. The time resolution in the plastic scintillator TOF barrel region is 68~ps, while
that in the end cap region was 110~ps. The end cap TOF
system was upgraded in 2015 using multigap resistive plate chamber
technology, providing a time resolution of
60~ps, which benefits $80\%$ of the data used in this analysis~\cite{etofa,etofb,etofc}.

Monte Carlo (MC) simulated data samples produced with a {\sc
geant4}-based~\cite{geant4} software package, which
includes the geometric description of the BESIII detector and the
detector response, are used to determine detection efficiencies
and to estimate backgrounds. The simulation models the beam
energy spread and ISR in the $e^+e^-$
annihilations with the generator {\sc
kkmc}~\cite{ref:kkmca,ref:kkmcb}.~All particle decays are modelled with {\sc
evtgen}~\cite{ref:evtgen,ref:evtgena} using branching fractions
either taken from the
Particle Data Group~(PDG)~\cite{PDG2018}, when available,
or otherwise estimated with {\sc lundcharm}~\cite{ref:lundcharma,ref:lundcharmb}. Final state radiation~(FSR)
from charged final state particles is incorporated using the {\sc
photos} package~\cite{ref:photos}. The inclusive MC sample includes the production of open charm
processes, the ISR production of vector charmonium(-like) states,
and the continuum processes incorporated in {\sc
kkmc}~\cite{ref:kkmca,ref:kkmcb}. The signal process is not included in the inclusive MC samples.~The inclusive MC samples, after removing processes including $J/\psi \to l^{+}l^{-}(l=e,\mu)$ in the final state, are denoted as “Bkg-Inc” in the following discussion.

The signal MC sample (denoted as PHSP MC) of $e^{+}e^{-}\to K^{0} K^-\pi^+ J/\psi+c.c., ~J/\psi \to l^{+}l^{-}$,~$K^{0} \to K_{S}^{0}~{\rm or}~K_{L}^{0}$, is simulated at each c.m.~energy point to determine the detection efficiency.~In the signal MC simulation, the dressed cross section line shape of $e^{+}e^{-} \to K^{0}K^-\pi^+ J/\psi+c.c.$ measured in this study is used as input to calculate the ISR correction factor.~To estimate
the potential background contribution, additional MC samples (denoted
as Bkg-Exc) of $e^{+}e^{-}\to \pi\pi\psi(3686),~K^{+}K^{-}J/\psi,~K_{S}^{0}K_{S}^{0}J/\psi,~\omega \chi_{c1,2},~\phi \chi_{c0,1,2},~\pi^{+}\pi^{-}X(3823)$,\\$\eta^{(')} J/\psi$,~$\psi(3686)\to{\rm anything}+J/\psi$,~$X(3823)\to\gamma\chi_{c1}$,~$\chi_{c0,1,2}\to\gamma J/\psi$,~$\omega\to\pi^{+}\pi^{-}\pi^{0}$,~$\phi(K_{S}^{0},\\\eta^{(')}) \to {\rm anything}$,~$\pi^{0}\to\gamma\gamma$,~and $J/\psi\to l^{+}l^{-}$ are generated, as these processes have similar final states with the signal processes. The cross sections of these processes are taken from the previous BESIII measurements~\cite{psi_3686_3,depth,kkjpsi,ksksjpsi,omegachic12,phichic0,phichic12,pipix3823,etajpsi}.~The Bkg-Inc and Bkg-Exc samples are used to study the sources of the background contributions and optimize the event selection criteria. 

\section{Event reconstruction and selection}
To improve the detection efficiencies, both full and partial reconstruction methods are employed.~The process $e^{+}e^{-} \to K^{0}_{L}K^-\pi^+ J/\psi+c.c.$ is reconstructed using the ``missing a $K^{0}_{L}$'' method, with the corresponding final states being $K^{\pm}\pi^{\mp}l^{+}l^{-}$. The decay chain $e^{+}e^{-} \to K^{0}_{S}K^-\pi^+ J/\psi+c.c.$ is reconstructed using ``full reconstruction'', ``missing a $K^{\pm}$ or $\pi^{\pm}$'', and ``missing a $K^{0}_{S}$'' methods.~These result in the final states of $K_{S}^{0}(\to \pi^{+}\pi^{-})K^{\pm}\pi^{\mp}l^{+}l^{-}$, $K_{S}^{0}(\to \pi^{+}\pi^{-})\pi^{\pm}l^{+}l^{-}~{\rm or}~K_{S}^{0}(\to \pi^{+}\pi^{-})K^{\pm}l^{+}l^{-}$, and $K^{\pm}\pi^{\mp}l^{+}l^{-}$, respectively. In the full text, ``missing a $K^{0}(\bar{K}^{0})$'' subsumes the methods of missing a $K^{0}_{L}$ or a $K^{0}_{S}$, as they share the same final state. The following criteria will be used to identify the candidate events.

%In this analysis, the cascade decays $e^{+}e^{-} \to K^{0}_{S}K^-\pi^+ J/\psi+c.c.$ 
\subsection{Event reconstruction}
  
Charged tracks detected in the MDC are required to be within a polar angle ($\theta$) range of $|\rm{cos\theta}|<0.93$, where $\theta$ is defined with respect to the $z$-axis,
which is the symmetry axis of the MDC. For charged tracks not originating from $K_S^0$ decays, the distance of closest approach to the interaction point (IP) must be less than 10\,cm
along the $z$-axis, $|V_{z}|$, and less than 1\,cm in the transverse plane, $|V_{xy}|$. Particle identification~(PID) for charged tracks combines measurements of the energy deposited in the MDC and the flight time in the TOF to form likelihoods $\mathcal{L}(h)~(h=K,\pi)$ for each hadron $h$ hypothesis.~Tracks are identified as kaons and pions by comparing the likelihoods for the kaon and pion hypotheses, $\mathcal{L}(K)>\mathcal{L}(\pi)$ and $\mathcal{L}(\pi)>\mathcal{L}(K)$, respectively.

The $K_{S}^0$ candidate is reconstructed from two oppositely charged tracks satisfying $P < 1.0~{\rm GeV}/c$ and $|V_{z}|<$ 20~cm, where $P$ represents the magnitude of the charged track's momentum. The two charged tracks are assigned as $\pi^+\pi^-$ without imposing further PID criteria. They are constrained to
originate from a common vertex and are required to have an invariant mass $M_{\pi^{+}\pi^{-}}$ satisfying $|M_{\pi^{+}\pi^{-}} - m_{K_{S}^{0}}|<$ 12~MeV$/c^{2}$, where $m_{K_{S}^{0}}$ is the $K^0_{S}$ nominal mass~\cite{PDG2018}. The
decay length of the $K^0_S$ candidate is required to be greater than twice the vertex resolution. If more than one $K_{S}^{0}$ are reconstructed in an event, the one with the smallest $\chi^{2}$ is kept as the final $K_{S}^{0}$ candidate, where $\chi^{2}$ is derived from the second vertex fit of $\pi^{+}\pi^{-}$.

The other two charged tracks with momenta greater than
$1.0~{\rm GeV}/c$ and opposite charges are identified as the
lepton pair from the $J/\psi$ decay, while the remaining charged tracks
are regarded as $K^{
\pm}$ or $\pi^{\pm}$ according to the PID information.~Electrons and muons are discriminated by requiring their deposited energies in
the EMC (labeled as $E$(EMC)) to be greater than $1.0~{\rm GeV}$ for electrons and less than $0.4~{\rm GeV}$ for muons.

The selected lepton pair, $K^{\pm}$, $\pi^{\pm}$, and $K^{0}_{S}$ candidates in the event are combined to reconstruct the signal process.~All charged tracks detected in the final states are used to perform a vertex fit to ensure they originate from the same vertex,~and the $\chi^{2}$ value of this fit is required to be less than 200. %If a $K_{S}^{0}$ is successfully reconstructed in the final state, the virtual particle from vertex fit of $\pi^{+}\pi^{-}$ will be used in the vertex fit.

\subsection{Event selection}
Depending on the number of $K^{\pm}$, $\pi^{\pm}$, and $K_{S}^{0}$, the selection methods are divided into three categories: ``full reconstruction'', ``missing a $K^{\pm}$ or $\pi^{\pm}$'', and ``missing a $K^{0}$ ($\bar{K}^{0})$''.~The selection criteria for each method are individually optimized using the Figure-of-Merit (FOM) defined as $\varepsilon/(5/2+\sqrt{B})$~\cite{fom}, where $\varepsilon$ is the detection efficiency of signal process, and $B$ is the number of background events estimated by using the Bkg-Inc and Bkg-Exc samples (labelled as Bkg). The Bkg samples from various processes are normalized according to the corresponding cross sections and luminosities of data samples.~Due to the limited statistics, the MC simulations from all the data samples are summed up in the optimization.  %The background MC samples from each data samples are then combined according to the luminosity of the data sample. 
%These definitions are always implied in this chapter. 
\subsubsection{Full reconstruction}
If all the particles in the final state can be successfully reconstructed, the event is selected using the ``full reconstruction'' method. A four-constraint (4C) kinematic fit is performed to ensure both energy and momentum conservation throughout the entire process, with a requirement of $\chi_{\rm 4C}^2<200$.~To reduce the background processes due to $\mu$/$\pi$ misidentification in the $\mu^{+}\mu^{-}$ channel, the hit depth in MUC (Depth) is required to satisfy (${\rm Depth}_{\mu^{+}}+{\rm Depth}_{\mu^{-}}$)~$>$~75 $\rm cm$. After imposing these requirements, the invariant mass distribution of $l^{+}l^{-}$ ($M_{l^{+}l^{-}}$) is shown in Fig.~\ref{fig:m_jpsi} (left), where a clear $J/\psi$ signal can be seen in data. The number of retained background events estimated by using the Bkg-Inc and Bkg-Exc samples is $0.5\pm0.7$.
\subsubsection{Missing a $K^{\pm}$ or $\pi^{\pm}$}
If the event contains a $\pi^{\pm}$ or $K^{\pm}$, at least one $K_S^0$, and a pair of leptons in the final state, the ``missing a $K^{\pm}$ or $\pi^{\pm}$'' method will be utilized.
In this method, a one-constraint (1C) kinematic fit is applied, where the four-momenta of $l^{+},~l^{-},~K^{\pm}~{\rm or}~\pi^{\pm}$, $K_{S}^{0}$ and a missing charged track are constrained to the initial four-momentum. The missing track's three-momentum is unknown, but its mass is known.~If the reconstructed track is identified as $\pi^{\pm}$, then the missing one is assumed to be a $K^{\mp}$, and vice versa.~In the $e^{+}e^{-}$ mode, a two-dimensional optimization is performed on the distributions of $E$(EMC) and $\chi^{2}_{\rm 1C}$, and the requirements are set to be $\chi^{2}_{\rm 1C} < 30$ and $E{\rm(EMC)} > 1.25~{\rm GeV}$ for at least one lepton to suppress background events.~For the $\mu^{+}\mu^{-}$ mode, the requirements of $\chi^{2}_{\rm 1C}<20$ and (${\rm Depth}_{\mu^{+}}+{\rm Depth}_{\mu^{-}}$) $>$ 75 $\rm cm$ are imposed, according to the optimization.~After applying the aforementioned selection criteria, the invariant mass distribution of $l^{+}l^{-}$ of the accepted candidates ($M_{l^+l^-}$) is depicted in Fig.~\ref{fig:m_jpsi} (middle). The total number of background events estimated from the Bkg-Inc and Bkg-Exc samples amounts to $6.6\pm2.6$.

\subsubsection{Missing a $K^{0}(\bar{K}^{0})$}
If the $K^{0}(\bar{K}^{0})$ oscillates to $K_{L}$, or 
$K_{S}^{0}$ but failed to be reconstructed using the $\pi^{+}\pi^{-}$ decay channel, and the numbers of $K^{\pm}$, $\pi^{\pm}$, and lepton pair are all greater than 0, the ``missing a $K^{0}$ ($\bar{K}^{0}$)'' method is applied.~In this method, a one-constraint (1C) kinematic fit is applied, where the four-momenta of $K^{\pm}$, $\pi^{\pm}$, $l^{+}$, $l^{-}$ and a missing $K^{0}$ ($\bar{K}^{0}$) are constrained to the initial four-momentum. The three-momentum of missing $K^{0}$ ($\bar{K}^{0}$) are unknown, but its mass is constrained at the nominal $K^{0}$ ($\bar{K}^{0}$) mass~\cite{PDG2018}.~If there are more than one $K^{\pm}\pi^{\mp}$ combination in one event, the one yielding the smallest $\chi^{2}_{\rm 1C}$ from the kinematic fit is preserved for subsequent analysis.~To remove the background contributions from the QED processes ($e^{+}e^{-}\to e^{+}e^{-}(n\gamma)$,~$e^{+}e^{-}\to \mu^{+}\mu^{-}(n\gamma)$,~$e^{+}e^{-}\to n\gamma$), the opening angles of $e^{\pm}K^{\mp}$, $e^{\pm}\pi^{\mp}$ and $K^{\pm}\pi^{\mp}$ are required to satisfy $|{\rm cos}\theta_{e^{\pm}K^{\mp},e^{\pm}\pi^{\mp},K^{\pm}\pi^{\mp}}| < 0.98$ for the $e^{+}e^{-}$ mode. In addition, the criteria $\chi^{2}_{\rm 1C} < 30$ and $E{\rm(EMC)} > 1.3~{\rm GeV}$ for at least one lepton are applied. For the $\mu^{+}\mu^{-}$ mode, the requirements $\chi^{2}_{\rm 1C}<10$ and (${\rm Depth}_{\mu^{+}}+{\rm Depth}_{\mu^{-}}$) $>$ 80 $\rm cm$ are used to suppress background events.~After the above selections, the distribution of $M_{l^{+}l^{-}}$ of the accepted candidates is shown in Fig.~\ref{fig:m_jpsi} (right), where the total number of background events estimated by using the Bkg-Inc and Bkg-Exc samples is $17.5\pm4.2$.

After all the event selection, the total number of background events estimated by using Bkg samples, summed over all the energy points and all the selection categories, is $24.5\pm5.0$. The fraction of the Bkg-Exc background events is about 66\%, and that from the Bkg-Inc background events is about 34\%. The Bkg-Inc background events are dominated by hadron processes and smoothly distributed in the $M_{l^{+}l^{-}}$ distribution. The $M_{l^{+}l^{-}}$ distributions from data and Bkg are also displayed for each energy point, as shown in Fig.~\ref{fig:mean1}.~In the following sections, ``full reconstruction'', ``missing a $K^{\pm}$ or $\pi^{\pm}$'', and ``missing a $K^{0}(\bar{K}^{0})$'' selection methods are labeled as method I, method II, and method III, respectively.
\section{Signal yields and cross sections}
\subsection{Signal yield extraction}
\begin{figure}[htp]
 \centering
\includegraphics[width=0.325\textwidth]{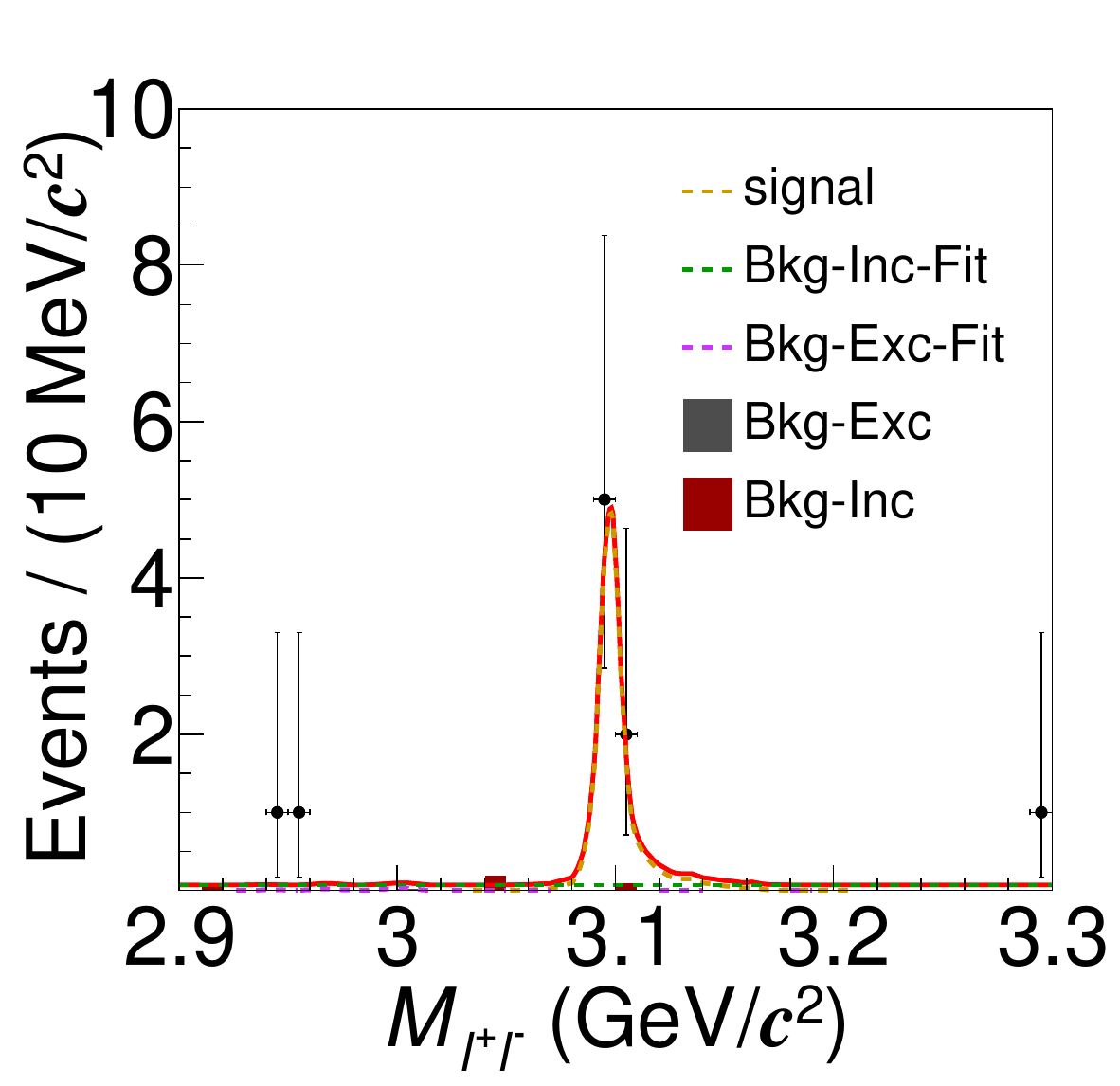}
\includegraphics[width=0.325\textwidth]{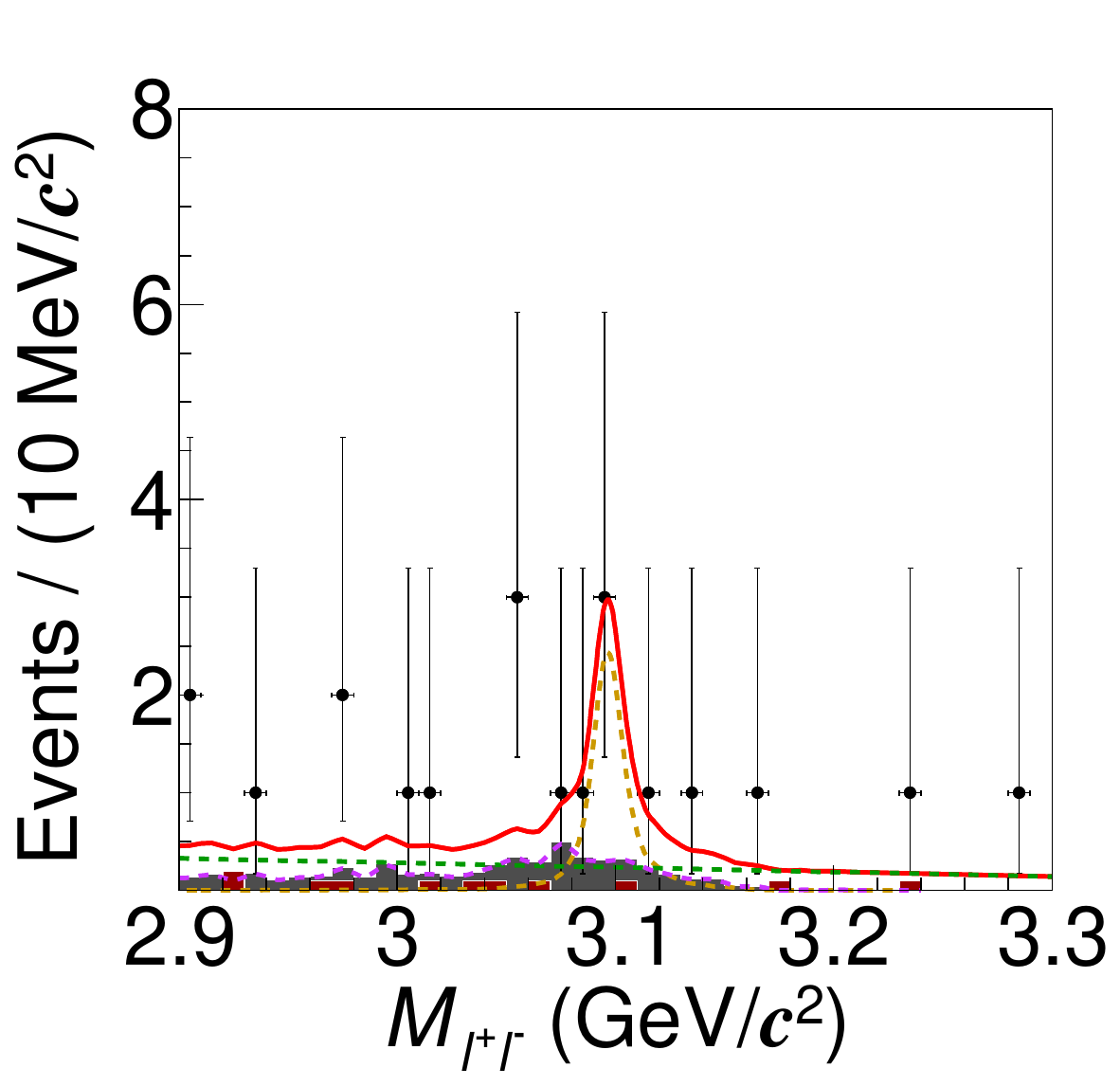}
\includegraphics[width=0.325\textwidth]{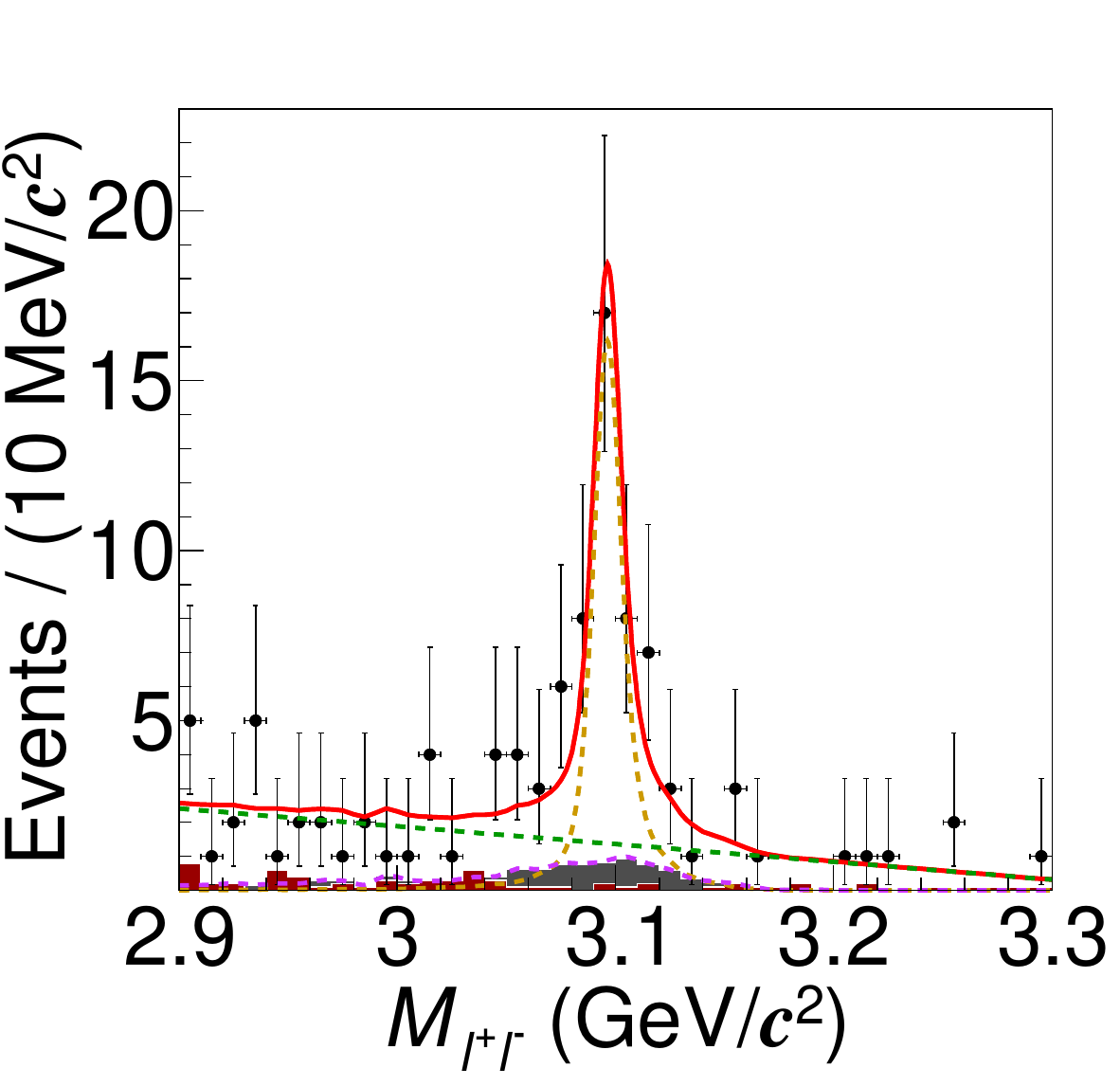}
\clearpage
 \caption{The $M_{l^{+}l^{-}}$ distributions from ``full reconstruction'' (left), ``missing a $K^{\pm}$ or $\pi^{\pm}$'' (middle), and ``missing a $K^{0}(\bar{K}^{0})$'' (right) methods.~The red and gray histograms represent the background contributions estimated from the Bkg-Inc and Bkg-Exc samples, respectively.~The dots with error bars are the experimental data (summed over all data samples) and the red solid lines are the best fit result from a simultaneous fit.
 }
\label{fig:m_jpsi}
\end{figure}

A simultaneous unbinned
maximum likelihood fit is performed to the $M_{l^{+}l^{-}}$
distributions from the three selection methods for all the data samples, as shown in Fig.~\ref{fig:m_jpsi}. The signal probability density function in the fit is represented by the MC simulated $J/\psi$ shape convolved with a Gaussian function, which accounts for the difference in mass resolution between data and MC simulation. Here, the signal MC shape is extracted from combined MC samples, where the signal MC simulations at each energy point are scaled according to the product of the luminosity and the detection efficiency, and then summed up. The parameters of the Gaussian function are obtained from the control sample $e^{+}e^{-}\to \pi^{+}\pi^{-}\psi(3686)$, $\psi(3686)\to \pi^{+}\pi^{-} J/\psi, J/\psi\to l^{+}l^{-}$, since the final state is similar to our signal process but with a larger statistics. In the simultaneous fit, the number of signal events is a free parameter and the ratios of the signal yield from three selection methods are fixed according to the detection efficiencies.~The background in the fit consists of two components. One is described by the Bkg-Exc MC shape, which is extracted from the dedicated MC simulations and normalized, with the number of events fixed to the estimated value.~The remaining background component is described by a constant for the method I and a linear function for the method II and III, where the numbers of events are free parameters. The best fit result is shown in Fig.~\ref{fig:m_jpsi},~and the total signal yield obtained from the fit is $45.3\pm7.8$.~The statistical significance is evaluated using the formula $\sqrt{2\ln(L_{\rm max}/L_{\rm 0})]/(\Delta\mathrm{n.d.f}=1)}$, where 
$\ln L_{\rm max}$ and $\ln L_{0}$ are the logarithmic likelihood values with and without the signal component in the 
fits, and $\Delta\mathrm{n.d.f}$ denotes the change in the number of degrees of freedom.~The signal significance is estimated to be $9.4 \sigma$.

\begin{figure}[h!]
 \centering
\includegraphics[scale=0.77]{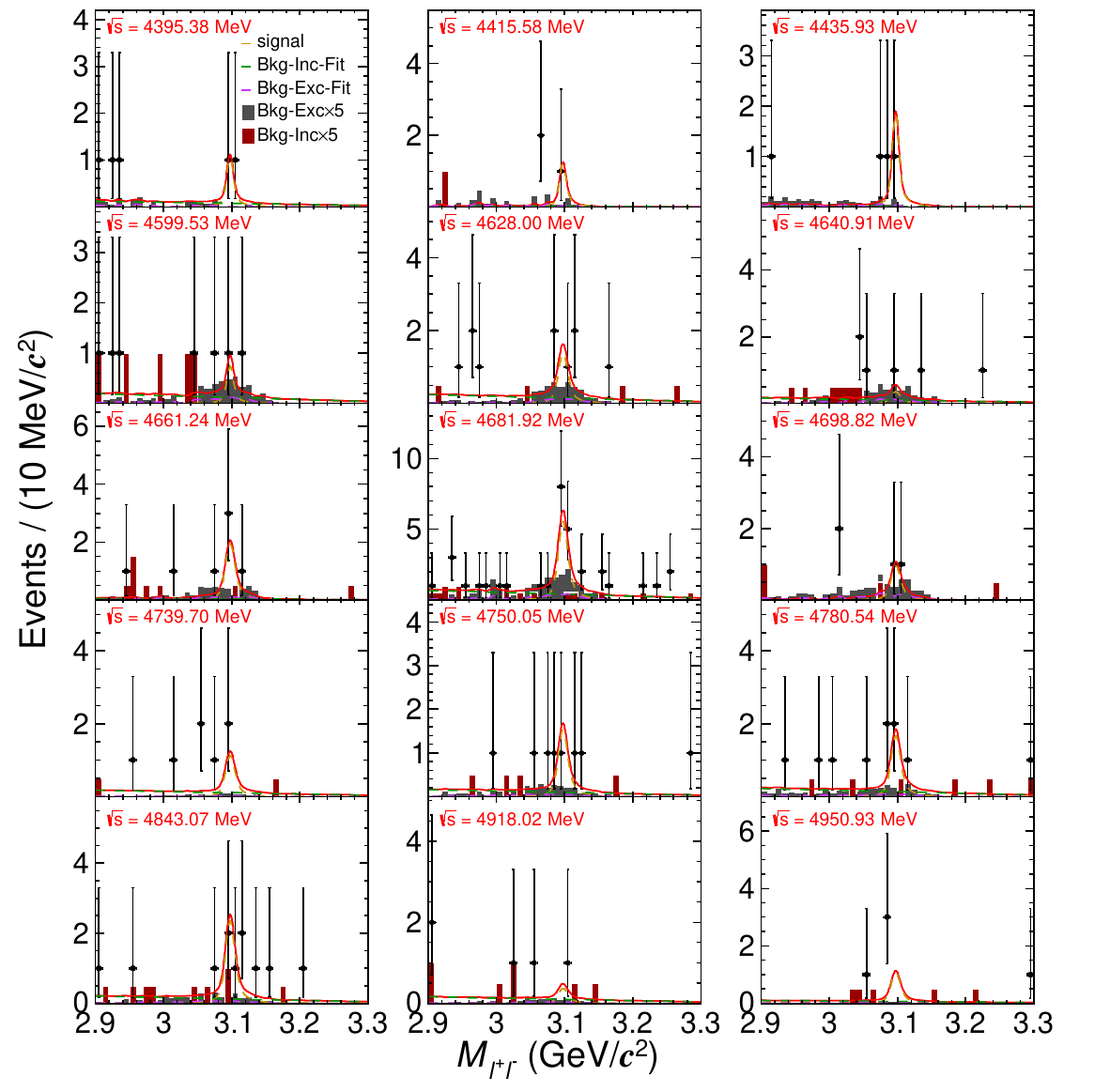}
 \caption{The fit results for data samples obtained by combining the three selection methods at different energy points. The red and gray histograms represent the background contributions estimated from the Bkg-Inc and Bkg-Exc samples, respectively. The dots with error bars are the data and the red solid lines are the best fit results.~The Bkg events are scaled up by five to provide a clearer delineation of their contribution.}
\label{fig:mean1}
\end{figure}

To extract the signal yield for each individual data sample, the $M_{l^{+}l^{-}}$ distributions from the three methods are combined and fitted due to limited statistics. In each fit, the signal shape is described by the corresponding signal MC shape convolved with a Gaussian function, where the mean and sigma values of the Gaussian function are taken from the control sample of $e^{+}e^{-}\to \pi^{+}\pi^{-}\psi(3686)$, $\psi(3686)\to \pi^{+}\pi^{-} J/\psi, J/\psi\to l^{+}l^{-}$.~The smooth background contributions are modeled by a linear function, with parameters fixed to the values obtained from fitting the $M_{l^{+}l^{-}}$ distribution of all data samples that passed the three sets of selections.~The numbers of signal and background events are free parameters.~Other background contributions estimated by the Bkg-Exc samples are included and fixed for both shape and yield in the fit.~The best fit results are shown in Fig.~\ref{fig:mean1}.~Due to the limited numbers of events that survived the final selection for the data samples taken at $\sqrt{s} = 4467.06$, $4527.14$, $4574.50$, and $4611.86~{\rm MeV}$, the fits on these samples are not feasible.~Instead, the number of signal events is calculated using $N_{\rm obs} - N_{\rm Bkg}$. Here, $N_{\rm obs}$ denotes the number of observed events within the $J/\psi$ signal region $(3.08,~3.12)~{\rm GeV}/c^{2}$, and $N_{\mathrm{Bkg}}$ represents the total background events in the same region, which include both $J/\psi$ peaking-background and non-$J/\psi$ events. The peaking-background contribution is estimated using the Bkg-Exc samples within the $J/\psi$ signal region, while the non-$J/\psi$ background is evaluated from the $J/\psi$ sideband regions, defined as $(3.02,~3.06)~{\rm GeV}/c^{2}$ and $(3.14,~3.18)~{\rm GeV}/c^{2}$.~The contribution from the $J/\psi$ sideband regions is scaled by 0.5, since the $J/\psi$ sideband regions are two times as wide as the $J/\psi$ signal region.~The signal yields are summarized in Table~\ref{tab:6}.~The significance of the signal process for each data sample is estimated using the same method mentioned above, and the values are listed in Table~\ref{tab:6}.
\\ \indent The statistical significance of the signal process for most data samples is below $3\sigma$, the upper limit of the number of signal events ($N_{\rm sig}^{\rm UL}$) at $90\%$ confidence level (C.L.) for these data samples is determined using a Bayesian method~\cite{ref:Bayesianmethod}.
The $M_{l^+l^-}$ distribution is fitted with the number of signal events fixed from $0$ to $n$ to obtain a series of likelihood values $L$,~where $n$ represents an extremely large value.~The upper limit is determined by finding out the value of $N_{\rm sig}^{\rm UL}$ corresponding to $90\%$ of the likelihood profile with respect to the signal yield. The upper limits at $90\%$ C.L. are listed in Table~\ref{tab:6}. Assuming that the signal and background yields follow a Poisson distribution, the {\textsc{trolke}} package \cite{t1} in the {\textsc {cernroot}} framework \cite{t2} is used to determine the upper limit of the number of signal events at $90\%$ C.L. for the data samples taken at $\sqrt{s}~=~4467.06,~4527.14,~4574.50,~{\rm and}~4611.86~{\rm MeV}$.  

\begin{table}[htbp]
\centering
\scalebox{0.65}{
\begin{threeparttable}
\caption{The Born cross section for different energy points,~where the first and the second uncertainties are statistical and systematic, respectively.~``$\sigma_{\rm UL}^{\rm Born}$ with sys.'' stands for the upper limits of the cross section with systematic uncertainties.~``Significance'' represents the statistical significance. $\varepsilon_{K^{0}_{S}}$ and $\varepsilon_{K^{0}_{L}}$ represent the detection efficiency of the process $e^{+}e^{-} \to K^{0}_{S}K^-\pi^+ J/\psi+c.c.$  and $e^{+}e^{-} \to K^{0}_{L}K^-\pi^+ J/\psi+c.c.$, respectively. Given the limited statistics for the data samples taken at $\sqrt{s}~=~4467.06,~4527.14,~4574.50,~{\rm and}~4611.86~{\rm MeV}$, the statistical significance is not evaluated.~The signal significance exceeds $3\sigma$ for the data samples taken at $\sqrt{s}~=~4661.24,~4681.92,~{\rm and}~4843.07~{\rm MeV}$.}
\label{tab:6}
\begin{tabular}{c c c c c c c c c c}
\toprule
$\sqrt{s}$ (MeV) & $\varepsilon_{K^{0}_{S}}$~($\%$) & $\varepsilon_{K^{0}_{L}}$~($\%$) & $N_{\rm sig}$ & $N^{\rm UL}_{\rm sig}$ & $\frac{1+\delta^r}{|1-\Pi|^2}$ & $L_{\rm int}~({\rm pb}^{-1})$ & $\sigma^{\rm Born}$~(pb) & $\sigma_{\rm UL}^{\rm Born}$ with sys.~(pb) & Significance ($\sigma$)\\
\midrule
4395.38&$25.0\pm0.1$&$15.9\pm0.1$&$1.7^{+1.8}_{-1.1}$&$5.1$&$0.850$&$507.8$&$0.16^{+0.17}_{-0.11}\pm0.01$&0.5&1.9\\
4415.58&$26.4\pm0.1$&$17.0\pm0.1$&$1.7^{+2.2}_{-1.2}$&$5.6$&$0.804$&$1090.7$&$0.07^{+0.10}_{-0.05}\pm0.0$&0.3&1.9\\
4435.93&$27.5\pm0.1$&$18.7\pm0.1$&$2.7^{+2.1}_{-1.4}$&$6.5$&$0.830$&$569.9$&$0.20^{+0.16}_{-0.11}\pm0.01$&0.5&2.7\\
4467.06&$26.1\pm0.1$&$18.2\pm0.1$&$-0.0^{+1.1}_{-1.1}$&$2.0$&$0.810$&$111.1$&$-0.01^{+0.48}_{-0.48}\pm0.0$&0.8&$-$\\
4527.14&$27.5\pm0.1$&$20.2\pm0.1$&$-0.0^{+1.1}_{-1.1}$&$2.0$&$0.822$&$112.1$&$-0.01^{+0.44}_{-0.44}\pm0.0$&0.8&$-$\\
4574.50&$28.5\pm0.1$&$21.0\pm0.1$&$-0.5^{+1.2}_{-1.2}$&$1.6$&$0.829$&$48.9$&$-0.44^{+1.02}_{-0.96}\pm0.03$&1.3&$-$\\
4599.53&$31.9\pm0.1$&$24.0\pm0.1$&$1.2^{+1.9}_{-1.2}$&$5.1$&$0.834$&$586.9$&$0.07^{+0.12}_{-0.07}\pm0.01$&0.3&1.1\\
4611.86&$28.2\pm0.1$&$21.5\pm0.1$&$0.9^{+2.3}_{-1.4}$&$3.7$&$0.837$&$103.7$&$0.36^{+0.89}_{-0.55}\pm0.03$&1.4&$-$\\
4628.00&$31.9\pm0.1$&$23.9\pm0.1$&$2.9^{+2.7}_{-1.9}$&$7.7$&$0.840$&$521.5$&$0.20^{+0.18}_{-0.13}\pm0.02$&0.5&1.7\\
4640.91&$32.1\pm0.1$&$24.4\pm0.1$&$0.7^{+1.6}_{-0.9}$&$4.4$&$0.841$&$551.7$&$0.05^{+0.11}_{-0.06}\pm0.0$&0.3&0.7\\
4661.24&$32.4\pm0.1$&$24.7\pm0.1$&$4.3^{+2.7}_{-2.0}$&$-$&$0.844$&$529.4$&$0.28^{+0.18}_{-0.13}\pm0.02$&$-$&3.3\\
4681.92&$32.5\pm0.1$&$25.0\pm0.1$&$12.2^{+4.5}_{-3.7}$&$-$&$0.847$&$1667.4$&$0.25^{+0.09}_{-0.08}\pm0.02$&$-$&4.7\\
4698.82&$32.6\pm0.1$&$25.3\pm0.1$&$1.6^{+1.8}_{-1.2}$&$5.3$&$0.849$&$535.5$&$0.11^{+0.12}_{-0.07}\pm0.01$&0.3&1.8\\
4739.70&$33.2\pm0.1$&$25.7\pm0.1$&$2.4^{+2.3}_{-1.5}$&$6.9$&$0.853$&$163.9$&$0.50^{+0.48}_{-0.31}\pm0.03$&1.4&2.1\\
4750.05&$33.2\pm0.1$&$25.6\pm0.1$&$3.2^{+2.8}_{-2.1}$&$8.1$&$0.854$&$366.6$&$0.29^{+0.26}_{-0.19}\pm0.02$&0.7&1.9\\
4780.54&$33.3\pm0.1$&$25.8\pm0.1$&$3.6^{+2.7}_{-2.0}$&$8.4$&$0.858$&$511.5$&$0.23^{+0.17}_{-0.13}\pm0.02$&0.5&2.4\\
4843.07&$33.7\pm0.1$&$26.8\pm0.1$&$5.2^{+3.0}_{-2.3}$&$-$&$0.865$&$525.2$&$0.32^{+0.19}_{-0.14}\pm0.03$&$-$&3.1\\
4918.02&$33.8\pm0.1$&$26.9\pm0.1$&$0.8^{+1.5}_{-0.8}$&$4.1$&$0.873$&$207.8$&$0.12^{+0.23}_{-0.12}\pm0.01$&0.6&1.0\\
4950.93&$33.5\pm0.1$&$27.0\pm0.1$&$2.4^{+2.3}_{-1.7}$&$6.7$&$0.875$&$159.3$&$0.49^{+0.46}_{-0.33}\pm0.04$&1.3&1.5\\
\bottomrule
\end{tabular}
\end{threeparttable}
}
\end{table}

\subsection{Cross section calculation}
The Born cross section of the signal process is calculated as:
\begin{small}
\begin{equation}
    \sigma^{\rm Born}[\EE\to K^{0}K^-\pi^+ J/\psi+c.c.] = \frac{\sigma^{\rm dressed}}{\frac{1}{|1-\Pi|^2}} = \frac{N_{\rm sig}}{L_{\rm int}\cdot(1+\delta^{r})\cdot \frac{1}{|1-\Pi|^2}\cdot{\mathcal{B}(J/\psi \to l^{+}l^{-})}\cdot \varepsilon},
\end{equation}
\end{small}where $N_{\rm sig}$ is the signal yield, $L_{\rm int}$ is the integrated luminosity, $\varepsilon$ is the detection efficiency,~and $\mathcal{B}$ is the branching fraction. %The detection efficiency $\varepsilon_{K_{S}^0}$ and $\varepsilon_{K_{L}^{0}}$ are determined by using signal MC simulations of $\EE\to K_{S}^{0}K^-\pi^+ J/\psi+c.c.$ and $\EE\to K_{L}^{0}K^-\pi^+ J/\psi+c.c.$ containing 200K events each.
The value of the vacuum polarization (VP) factor, $\frac{1}{|1-\Pi|^2}$, is calculated using a Fortran package provided by Fred Jegerlehner~\cite{Jegerlehner:2011mw}, where the corrections from leptonic and hadronic VP are included.~The ISR correction factor, $(1+\delta^{r})$, is calculated following the procedure described in Ref.~\cite{weight}.~The cross section line shape measured from this study is used as input for the ISR correction factor calculation. 
The obtained Born cross section at each energy point is listed in Table~\ref{tab:6}.

\begin{figure}[htp]
  \centering
\includegraphics[width=0.5\textwidth]{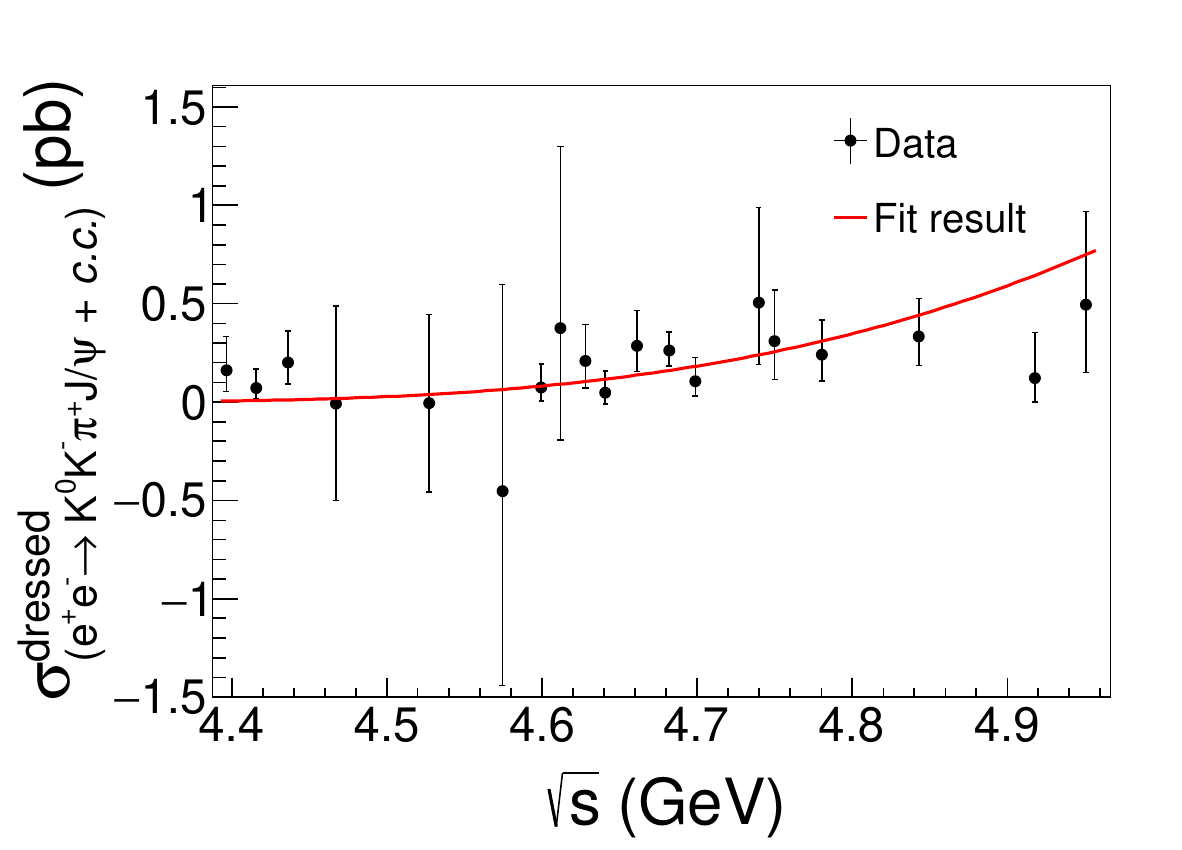}
  \caption{The dressed cross sections and fit results.}
  \label{fit_line}
\end{figure}

The dressed cross section line shape is fitted with a continuum 
amplitude:
\begin{small}
\begin{equation}
n_{0}\cdot PS(\sqrt{s})/s, 
\end{equation}
\end{small}where $n_{0}$ is free parameter, and $PS(\sqrt{s})$ is the four-body decay $[R(\sqrt{s}) \to K^{0}K^{-}\pi^{+} J/\psi~+~c.c.]$ phase space factor~\cite{PDG2018}. The result of the fit to the dressed cross section is shown in Fig.~\ref{fit_line}. 

\begin{figure}[H]
  \centering
\includegraphics[width=0.45\textwidth]{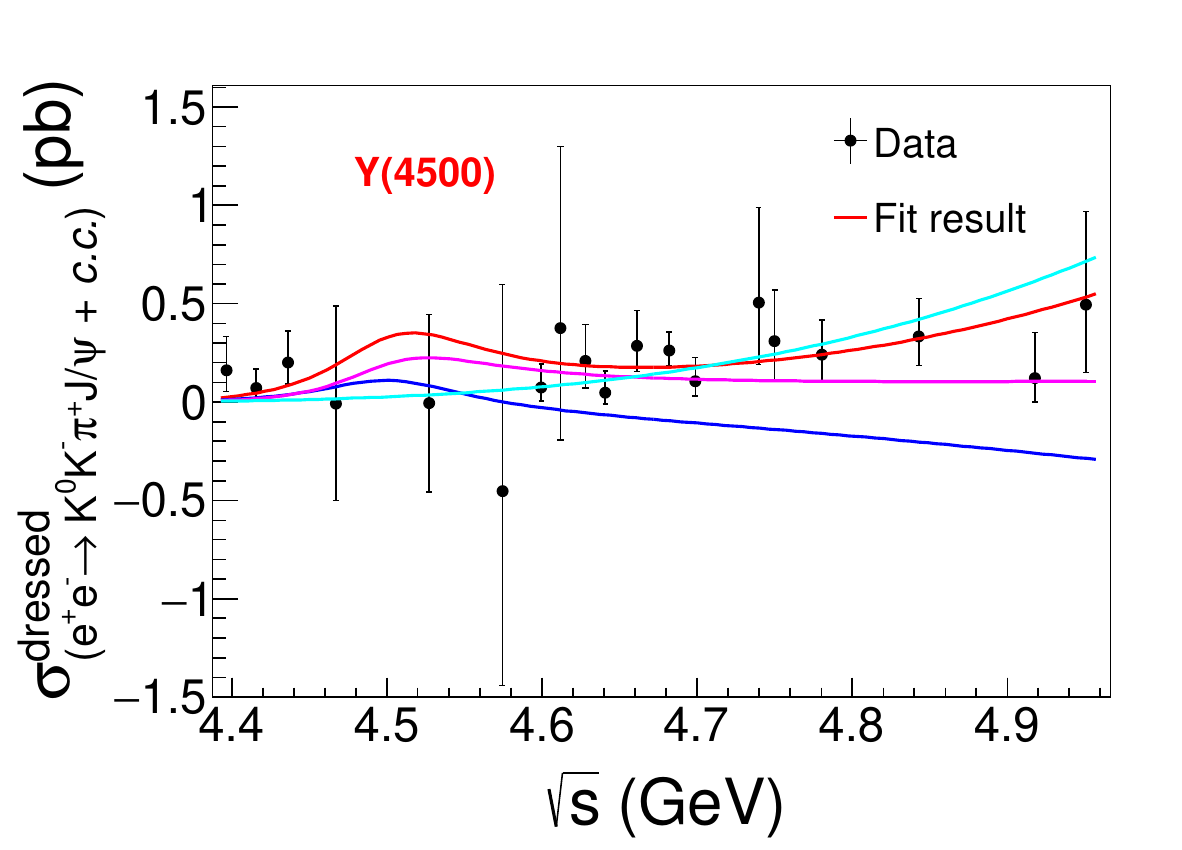}
\includegraphics[width=0.45\textwidth]{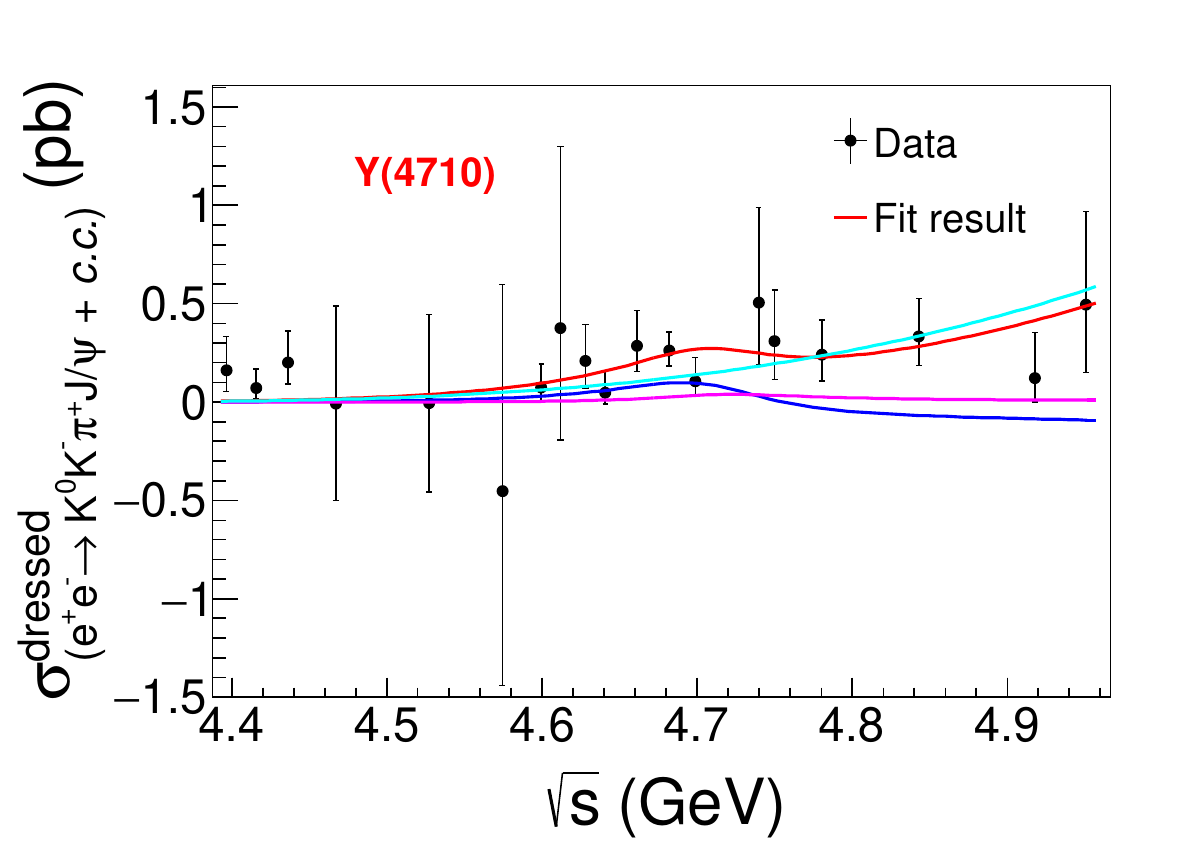}
\includegraphics[width=0.45\textwidth]{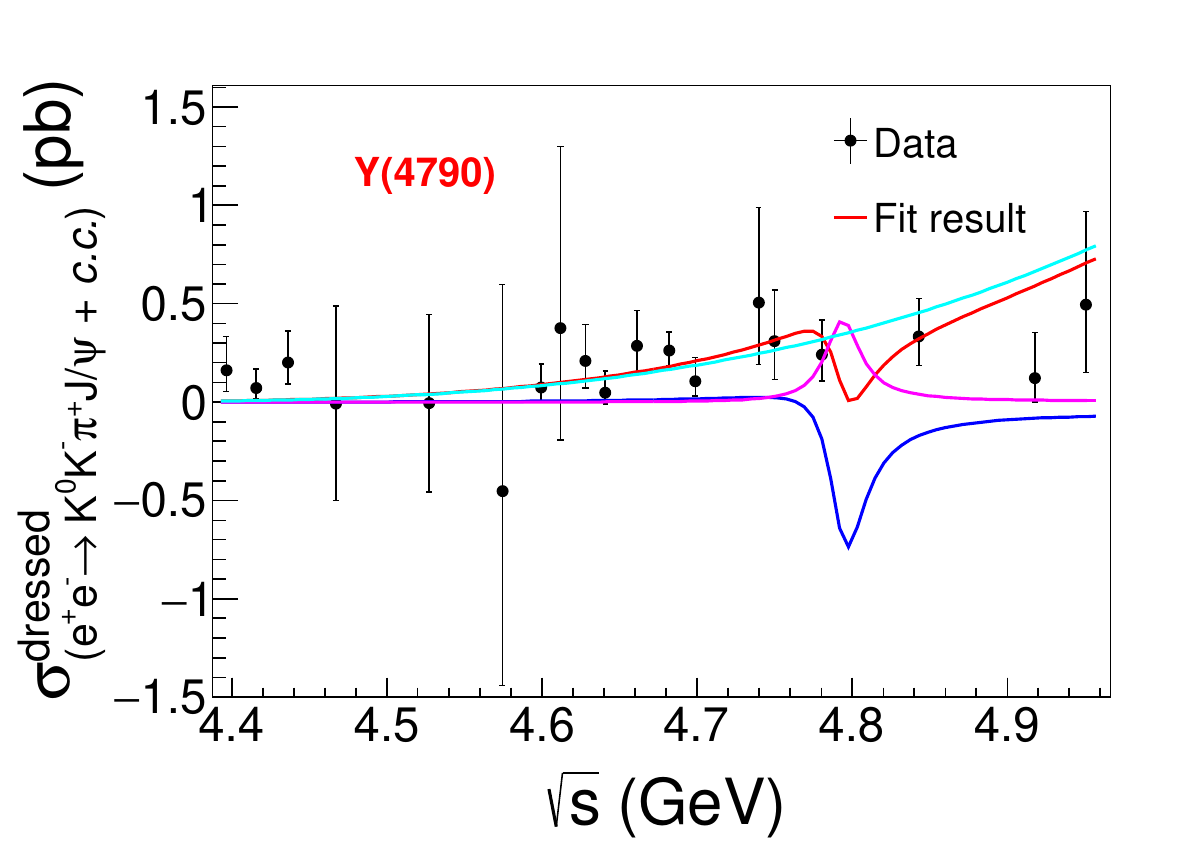}
\includegraphics[width=0.45\textwidth]{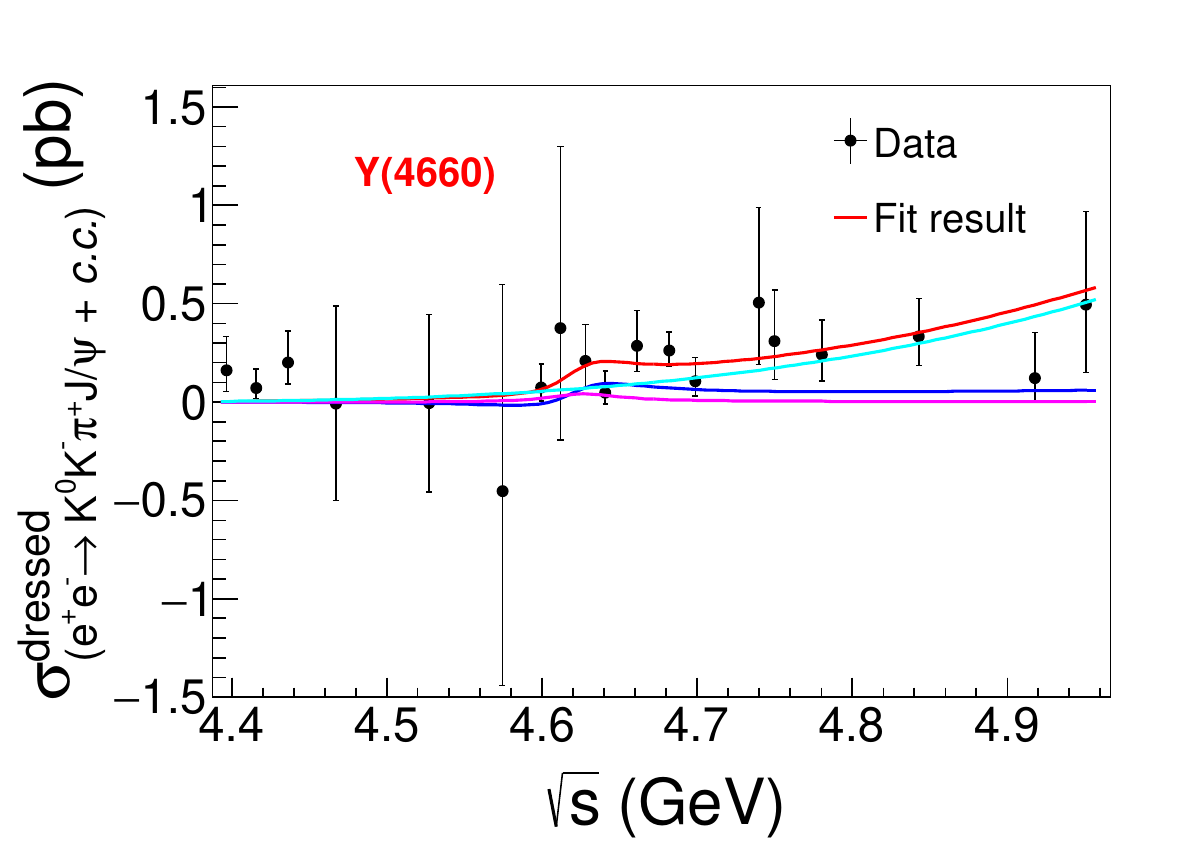}
  \caption{The dressed cross sections and fit results.~The contributions from the BW, interference, and continuum term are represented by the pink, blue, and cyan curves, respectively.}
  \label{fit_line_Y}
\end{figure}

The $Y$ states, reported in the processes $e^{+}e^{-}\to K^{+}K^{-}J/\psi$ and $e^{+}e^{-}\to D^{*+}_{s}D^{*-}_{s}$, are also searched for by including them in the fit. The cross section is fitted with a coherent sum of a Breit-Wigner (BW) function and the continuum amplitude:
\begin{small}
\begin{equation}
\left| \frac{\sqrt{PS(\sqrt{s})}\cdot(\sqrt{12\pi\Gamma_{ee}\mathcal{B}\Gamma})}{\sqrt{PS(M)}\cdot(s-M^{2}+iM\Gamma)} + \frac{n_{0}\cdot \sqrt{PS(\sqrt{s})}}{\sqrt{s}}\cdot e^{i\phi}\right|^{2}, 
\end{equation}
\end{small}where $M$, $\Gamma$, $\Gamma_{ee}$, and $\mathcal{B}$ are the mass, full width, electronic width, and the branching fraction to  $K^{0}K^{-}\pi^{+}J/\psi +c.c.$ final state of each $Y$ state, respectively.~The BW parameters are fixed to  that of $Y(4500)$~($M = 4499.4~{\rm MeV}/c^{\rm 2}$, $\Gamma = 124.0~{\rm MeV}$),  $Y(4710)$~($M = 4708.0~{\rm MeV}/c^{\rm 2}$, $\Gamma = 126.0~{\rm MeV}$), $Y(4790)$~($M = 4793.3~{\rm MeV}/c^{\rm 2}$, $\Gamma = 27.1~{\rm MeV}$), or $Y(4660)$~($M = 4623.0~{\rm MeV}/c^{\rm 2}$, $\Gamma = 55.0~{\rm MeV}$). These values are taken from the measurements of the processes $e^{+}e^{-}\to K^{+}K^{-}J/\psi$~\cite{depth}, $e^{+}e^{-}\to D^{*+}_{s}D^{*-}_{s}$~\cite{dssdss}, and PDG~\cite{PDG2018}. The fit results are shown in Fig.~\ref{fit_line_Y}, and the significance of the resonance contribution is less than $2\sigma$ in all fits. 
\section{Study of the possible intermediate states}
\begin{figure}[htp]
 \centering
 \subfigure{\includegraphics[scale=0.35]{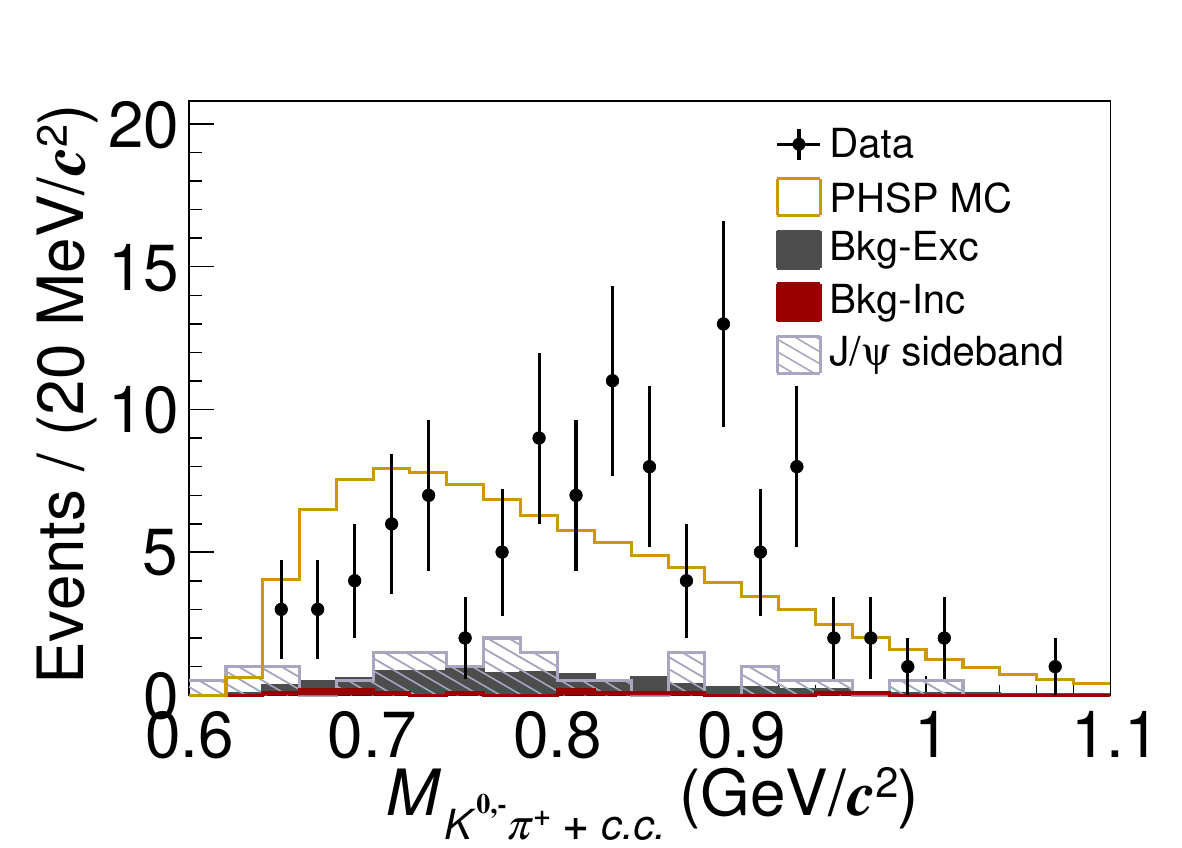}}
 \subfigure{\includegraphics[scale=0.35]{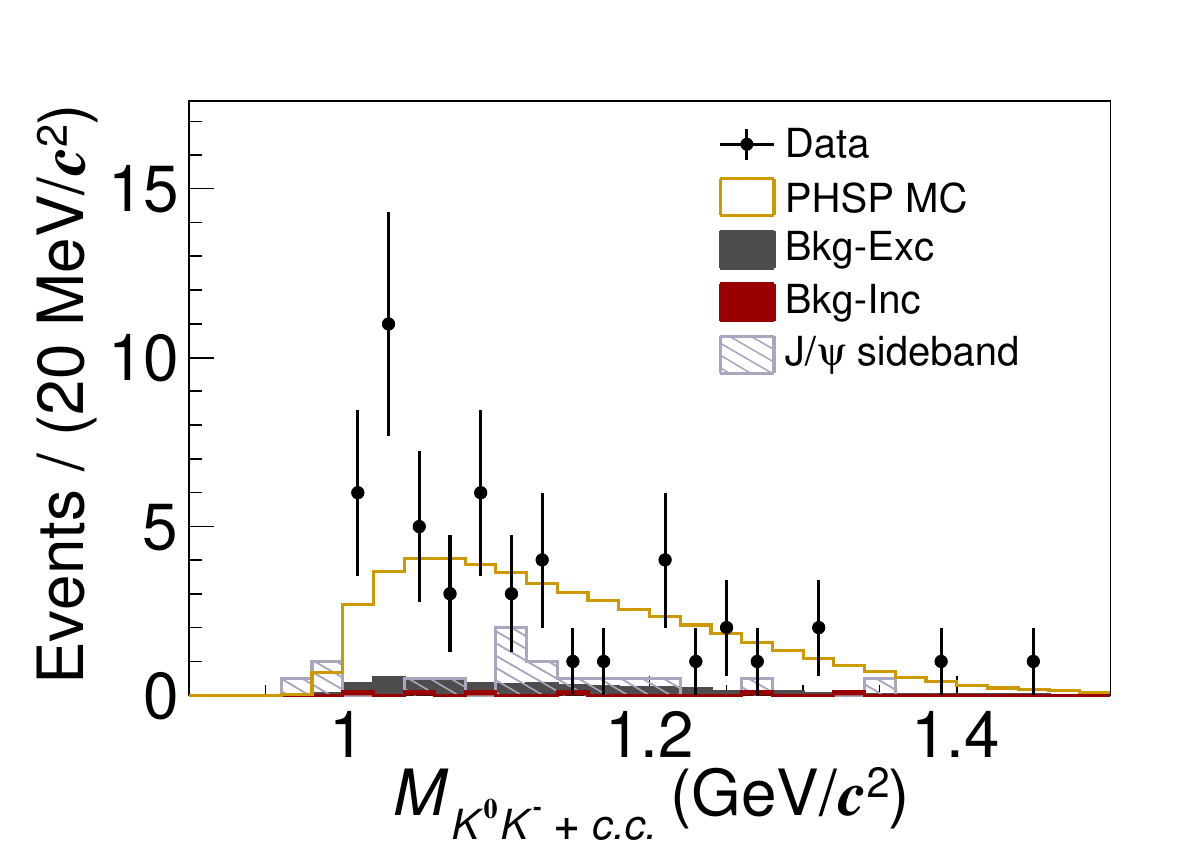}}
 \subfigure{\includegraphics[scale=0.35]{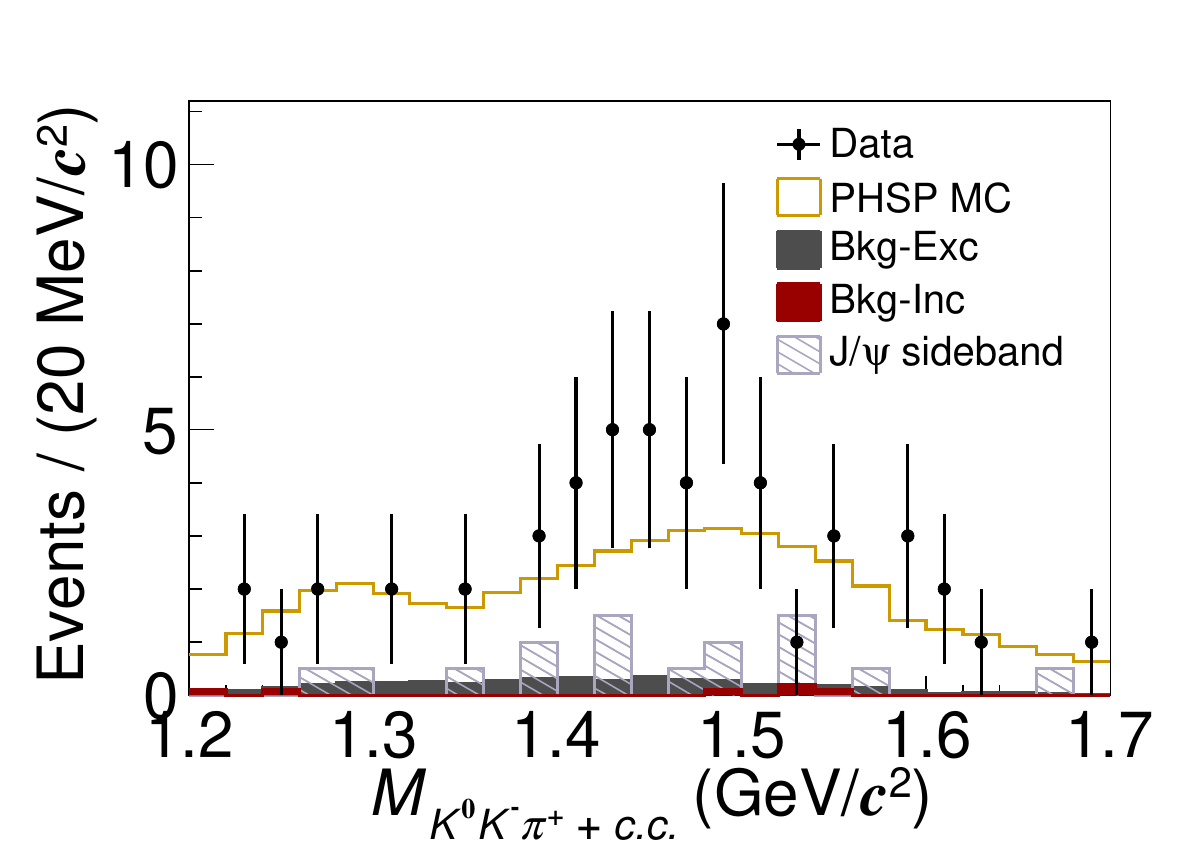}}
 \subfigure{\includegraphics[scale=0.35]{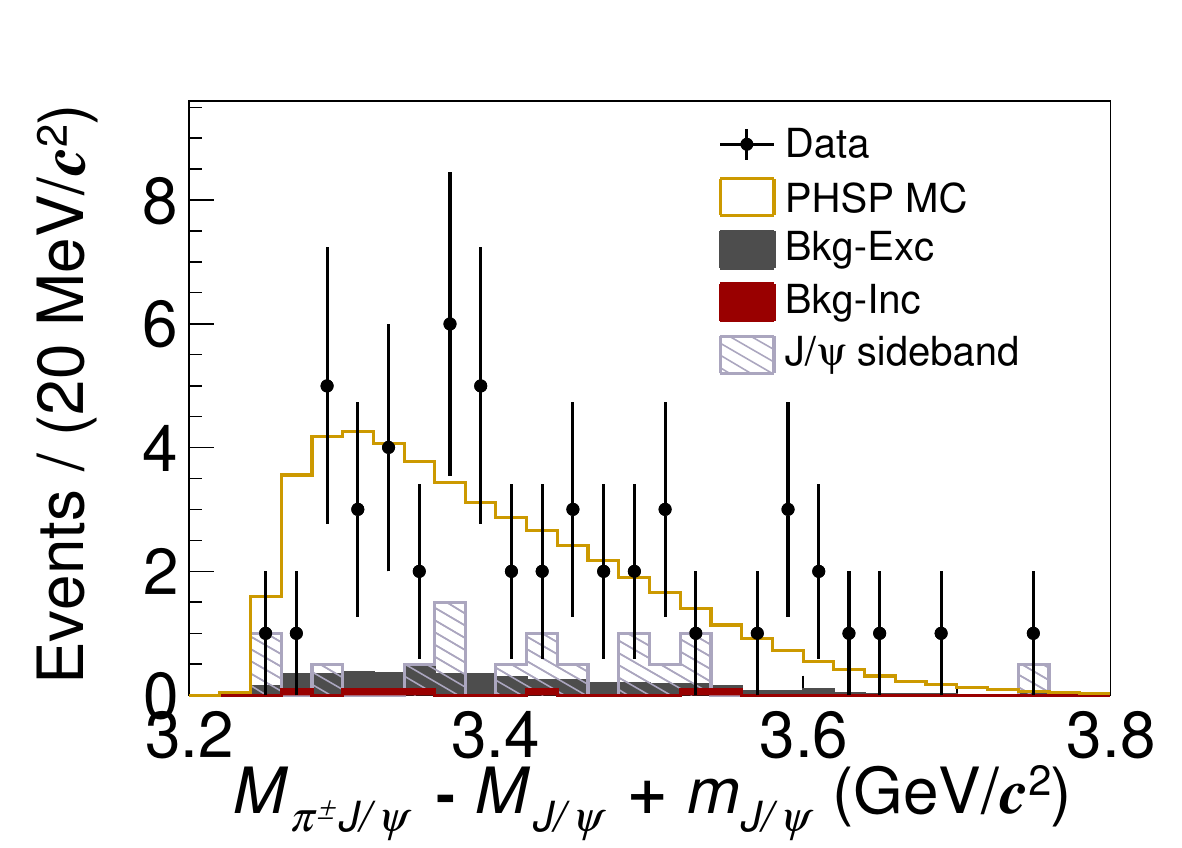}}
  \subfigure{\includegraphics[scale=0.35]{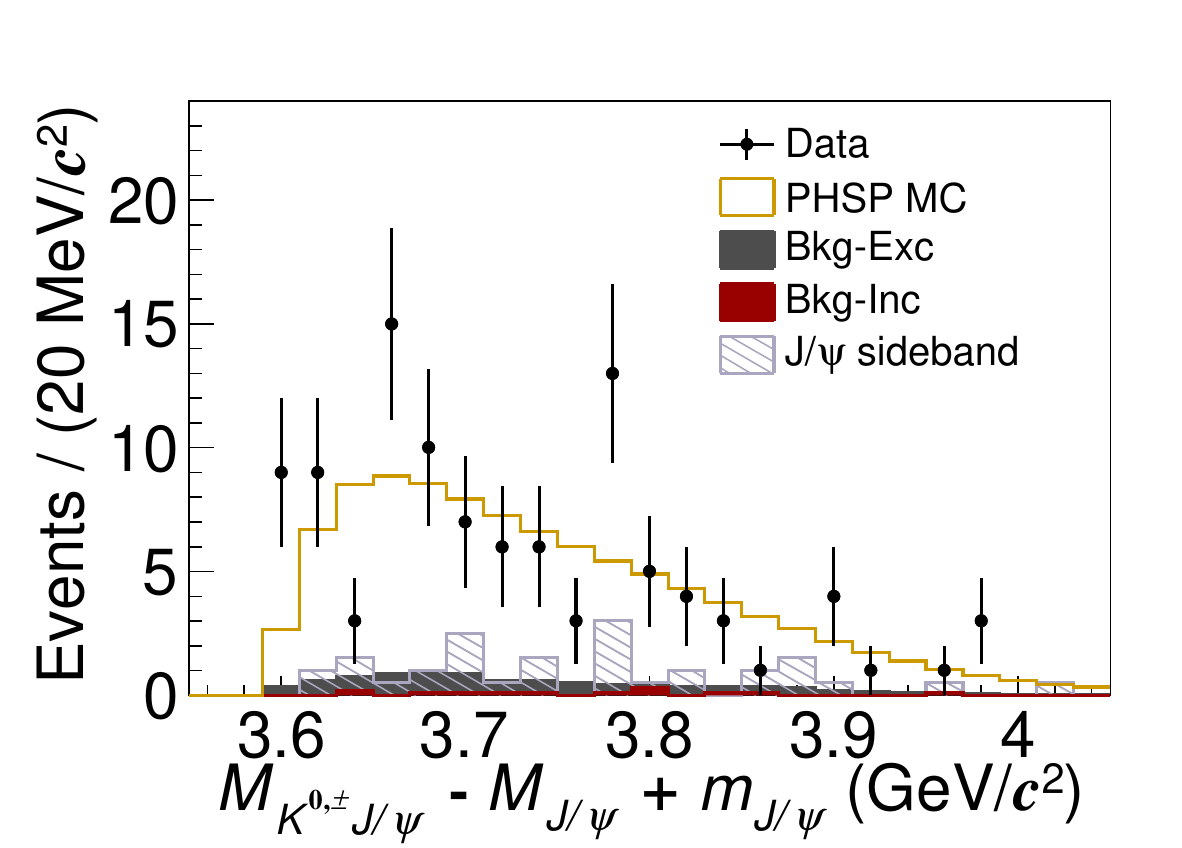}}
 \subfigure{\includegraphics[scale=0.35]{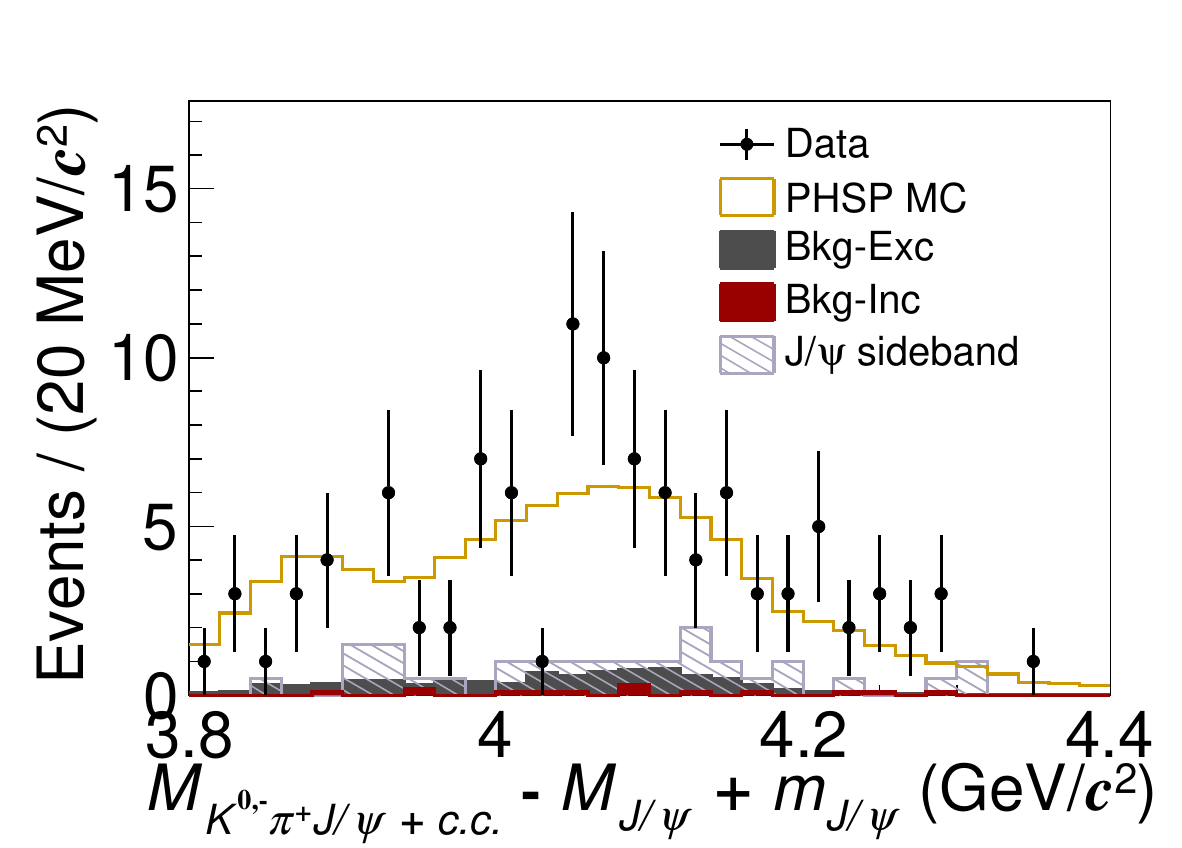}}
 \caption{The comparison of the invariant mass spectra of $K^{0,-}\pi^{+}+c.c.$, $K^{0}K^{-}+c.c.$, $K^{0}K^{-}\pi^{+}+c.c.$, $\pi^{\pm}J/\psi$, $K^{0,\pm}(\bar{K}^{0})J/\psi$, and $K^{0,-}\pi^{+}J/\psi+c.c.$ among data (dots with error bars, summing all data samples), PHSP MC, Bkg and $J/\psi$ sideband.~The signal MC samples at each energy point are scaled according to the product of the luminosity, the detection efficiency and the cross section, then added.}
\label{fig:inter}
\end{figure}

The selected events summed across all the data samples and selection methods are utilized to investigate the possible intermediate states in the $K^{0,-}\pi^{+}+c.c.$, $K^{0}K^{-}+c.c.$, $K^{0}K^{-}\pi^{+}+c.c.$, $\pi^{\pm}J/\psi$, $K^{0,\pm}(\bar{K}^{0})J/\psi$, $K^{0,-}\pi^{+}J/\psi+c.c.$ system, and the corresponding invariant mass distributions are illustrated in Fig.~\ref{fig:inter}. The contribution from the $J/\psi$ sideband regions is scaled by 0.5, a factor determined by fitting the $M_{l^{+}l^{-}}$ distribution from all data and calculating the ratio between the numbers of background events in the sideband and signal regions. The invariant mass of the $K^{0,-}\pi^{+}+c.c.$ ($M_{K^{0,-}\pi^{+}+c.c.}$) is utilized to investigate $K^{*}$ states, encompassing both the $K^{0}\pi^{+}+c.c.$ and $K^{\pm}\pi^{\mp}$ combinations, which results in each event being filled twice. The invariant mass of $K^{0}K^{-}+c.c.$ ($M_{K^{0}K^{-}+c.c.}$) is employed to study $a_{0}(980)$ states. The invariant mass of $K^{0}K^{-}\pi^{+}+c.c.$ ($M_{K^{0}K^{-}\pi^{+}+c.c.}$) is applied to analyze the excited $\eta$ states.
The invariant mass of $\pi^{\pm}J/\psi$ ($M_{\pi^{\pm}J/\psi}$) is used to study possible $Z_{c}$ states in the $\pi^{\pm}J/\psi$ system. The resolution of $M_{\pi^{\pm}J/\psi}$ is improved by utilizing the quantity $M_{\pi^{\pm}J/\psi}-M_{J/\psi}+m_{J/\psi}$, where $M_{J/\psi}$ represents the invariant mass of the lepton pair, and $m_{J/\psi}$ is the known mass as cited from the PDG~\cite{PDG2018}.~The invariant mass of the $K^{0,\pm}(\bar{K}^{0})J/\psi$ ($M_{K^{0,\pm}(\bar{K}^{0})J/\psi}$) and $K^{0,-}\pi^{+}J/\psi+c.c.$ ($M_{K^{0,-}\pi^{+}J/\psi+c.c.}$) are employed to search for the $Z_{cs}$ states.~To improve the mass resolution, the requirements of $M_{K^{0,\pm}(\bar{K}^{0})J/\psi}-M_{J/\psi}+m_{J/\psi}$ and $M_{K^{0,-}\pi^{+}J/\psi+c.c.}-M_{J/\psi}+m_{J/\psi}$ are used instead of $M_{K^{0,\pm}(\bar{K}^{0})J/\psi}$ and $M_{K^{0,-}\pi^{+}J/\psi+c.c.}$, respectively. 
The obtained distributions of these combinations for 
the signals in data are consistent with that of the four-body phase space, and
no obvious intermediate resonances are found as shown in Fig.~\ref{fig:inter}.
\section{Systematic uncertainties}
The systematic uncertainties on the Born cross section measurement are mainly caused by the integrated luminosity, tracking, PID, branching fraction, VP factor, ISR factor, kinematic fit, MUC hit depth, MC model, and fit procedure.~The systematic uncertainties from the opening angles of $e^{\pm}K^{\mp}$, $e^{\pm}\pi^{\mp}$ and $K^{\pm}\pi^{\mp}$ and $E{\rm(EMC)}$ requirements are found to be negligible, estimated by the Barlow test~\cite{sigma}.~The systematic uncertainties are summarized in Table~\ref{err}. The total systematic uncertainty is taken as a quadratic sum of each contribution. The systematic uncertainties in the determination of the upper limit of the cross section can be classified as either additive or multiplicative terms. 
\begin{table}[htp]
\centering
\scalebox{0.63}{
\begin{threeparttable}
\caption{Summary of the systematic uncertainties (in $\%$) for the Born cross section measurements, where the sources marked with * indicate additive uncertainties and the rest are multiplicative uncertainties. ``ISR-a'' and ``ISR-b'' refer to the systematic uncertainties arising from the parameterization form and the fit parameters of the cross section line shape, respectively. ``Bkg-a'' and ``Bkg-b'' denote the systematic uncertainties associated with the fixed Bkg-Exc background contribution and the background shapes, respectively. ``Kinematic fit'', ``Branching fraction'', and ``MUC hit depth'' are abbreviated as ``Kfit'', ``Br'', and ``MUC'', respectively.}
\label{err}
\begin{tabular}{ccccccccccccccc}
\toprule
$\sqrt{s}~({\rm MeV})$&$L_{\rm int}$&VP factor&Tracking&PID&Kfit&ISR-a&ISR-b&Br&MUC&MC model&Signal shape*&Bkg-a*&Bkg-b*&Total\\
\midrule
4395.38&1.0&0.5&2.3&0.3&0.3&4.7&0.1&0.4&0.9&$-$&0.9&3.1&0.4&6.3\\
4415.58&1.0&0.5&2.3&0.3&0.4&1.8&0.1&0.4&0.9&$-$&0.9&3.1&0.4&4.6\\
4435.93&1.0&0.5&2.3&0.3&0.3&1.7&0.1&0.4&0.9&$-$&0.9&3.1&0.4&4.6\\
4467.06&1.0&0.5&2.3&0.3&0.7&1.0&0.1&0.4&0.9&$-$&$-$&$-$&$-$&3.0\\
4527.14&1.0&0.5&2.3&0.3&0.7&1.4&0.1&0.4&0.9&0.2&$-$&$-$&$-$&3.1\\
4574.50&1.0&0.5&2.3&0.3&0.8&1.8&0.1&0.4&0.9&7.1&$-$&$-$&$-$&7.8\\
4599.53&1.0&0.5&2.3&0.3&0.5&3.2&0.1&0.4&0.8&5.9&0.9&3.1&0.4&8.0\\
4611.86&0.6&0.5&2.3&0.3&0.7&2.1&0.1&0.4&1.8&6.0&$-$&$-$&$-$&7.1\\
4628.00&0.6&0.5&2.3&0.3&0.4&3.6&0.1&0.4&2.7&5.4&0.9&3.1&0.4&8.1\\
4640.91&0.6&0.5&2.3&0.3&0.4&3.8&0.1&0.4&2.7&5.5&0.9&3.1&0.4&8.3\\
4661.24&0.6&0.5&2.3&0.3&0.4&3.9&0.1&0.4&2.4&5.4&0.9&3.1&0.4&8.2\\
4681.92&0.6&0.5&2.3&0.3&0.4&3.7&0.1&0.4&2.1&5.8&0.9&3.1&0.4&8.2\\
4698.82&0.6&0.5&2.3&0.3&0.4&3.0&0.1&0.4&1.4&5.4&0.9&3.1&0.4&7.5\\
4739.70&0.6&0.5&2.3&0.3&0.3&1.8&0.1&0.4&0.6&4.9&0.9&3.1&0.4&6.6\\
4750.05&0.6&0.5&2.3&0.3&0.4&3.3&0.1&0.4&2.1&5.0&0.9&3.1&0.4&7.6\\
4780.54&0.6&0.5&2.3&0.3&0.5&6.4&0.1&0.4&0.9&4.5&0.9&3.1&0.4&8.9\\
4843.07&0.6&0.5&2.3&0.3&0.5&7.4&0.1&0.4&2.5&4.1&0.9&3.1&0.4&9.8\\
4918.02&0.6&0.5&2.3&0.3&0.4&6.1&0.1&0.4&0.6&4.0&0.9&3.1&0.4&8.4\\
4950.93&0.6&0.5&2.3&0.3&0.4&5.5&0.1&0.4&0.5&3.1&0.9&3.1&0.4&7.6\\
\bottomrule
\end{tabular}
\end{threeparttable}
}
\end{table}
\begin{itemize}
\item \emph{Integrated luminosity.} The integrated luminosities are measured using Bhabha events ~\cite{lumb,BESIII:lum}, and the corresponding uncertainties are estimated to be $0.6\%$ for $\sqrt{s}>4.6~{\rm GeV}$ and $1.0\%$ for $\sqrt{s}<4.6~{\rm GeV}$.

\item \emph{Tracking.}~The systematic uncertainty of tracking is about $1.0\%$ per track~\cite{Ablikim:092009,Ablikim:2011kv,error_ppb2pi}.~Since events with one missing track are included using the above mentioned selection method, the combined relative systematic uncertainty is calculated with
\begin{equation}\label{eq:sys-combine}
\left| \frac{\varepsilon\cdot\varepsilon+2\cdot\varepsilon(1-\varepsilon)}{\varepsilon^{'}\cdot\varepsilon^{'}+2\cdot\varepsilon^{'}(1-\varepsilon^{'})}-1\right|+2\%,
\end{equation}
where $\varepsilon$ and $\varepsilon^{'}=\varepsilon\cdot(1\pm1\%)$ are the tracking efficiencies of MC simulation and data, respectively. The combined uncertainty is $2.3\%$.

\item \emph{PID.} This systematic uncertainty is calculated using
the same method as used for tracking, and the corresponding uncertainty is assigned to be $1.0\%$ per track~\cite{Ablikim:092009,Ablikim:2011kv,error_ppb2pi}.
\item \emph{Branching fraction.} The branching fractions of $J/\psi \to e^{+}e^{-}$ and $J/\psi \to \mu^{+}\mu^{-}$ are taken from the PDG~\cite{PDG2018}, and they contribute an uncertainty of $0.4\%$.

\item \emph{VP factor.} The uncertainty in the vacuum polarization factor is $0.5\%$, taken from a QED calculation~\cite{vac}.

\item \emph{ISR factor.}  To get the uncertainty from the parameterization form of the dressed cross section line shape, the fit formula is changed to $|\sqrt{PS(\sqrt{s})/PS(M)}\cdot(\sqrt{12\pi\Gamma_{ee}\mathcal{B}\Gamma})/(s-M^{2}+iM\Gamma)|^{2}$ and refit the dressed cross section. Here, $M$ and $\Gamma$ represent mass and total width, and are fixed to $4708.0~{\rm MeV}/c^{2}$ and $126.0~{\rm MeV}$ in the fit~\cite{depth}, respectively. Using the fitted curve to recalculate the ISR factor and detection efficiency, the difference from the nominal result is taken as systematic uncertainty. To get uncertainty from the fitted parameters, we change the parameter of $n_{0}$ with $\pm1\sigma$, to reiterate and determine $(1 + \delta^{r})\cdot \varepsilon$ values. The largest difference on the resultant distribution of $(1 + \delta^{r})\cdot \varepsilon$ is taken as the systematic uncertainty, and the values are listed in Table~\ref{err}. 
\item \emph{Kinematic fit.}
The helix parameters of the charged tracks are corrected in MC simulation to improve the agreement of the $\chi^{2}$ from the kinematic fit between data and simulation~\cite{helix}. The systematic uncertainty from the kinematic fit is assigned as the difference in the efficiencies with and without this correction. 

\item \emph{MUC hit depth.}
The uncertainty from the MUC response is studied with a control sample of $e^{+}e^{-}\to\mu^{+}\mu^{-}$ events. The difference in efficiency between the data and MC simulation is taken as the systematic uncertainty.
\item \emph{MC model.} Since the intermediate resonance states are indistinct, the four-body phase space model is used to obtain the nominal detection efficiency.~To estimate the systematic uncertainty of the MC model, the processes $e^{+}e^{-}\to K^{0}\bar{K}^{*0}(\to K^{-}\pi^{+}) J/\psi + c.c.$ and $e^{+}e^{-}\to K^{*+}(\to K^{0}\pi^{+})K^{-} J/\psi~+c.c.$ are generated.~The differences in efficiencies by MC samples with and without $K^{*}$ intermediate states are taken as the systematic uncertainties, as listed in Table~\ref{err}.
%(\to K^{\pm} \pi^{\mp}, K^{0} \pi^{\pm}+c.c.)
\item \emph{Fit procedure.}~To reduce statistical fluctuations, all data samples are combined to estimate the systematic uncertainty related to fit conditions. The fit procedure involves uncertainties from both the signal and background shapes.~The signal MC shape convolved with a Gaussian function is used to describe the signal shape in the nominal fit.~By varying the mean and standard deviation of the Gaussian function within one standard deviation and repeating the fit, the largest difference in the number of signal events is taken as the systematic uncertainty. For the background shapes, the parameters of the linear function are changed with $\pm1\sigma$, and the fit is repeated.~The largest difference in the number of signal events is considered as the systematic uncertainty due to the background shapes.~In the nominal fit, the number of Bkg-Exc events is fixed to $\sum L_{\rm int} \cdot \sigma_k \cdot \epsilon_k$, where $\sigma_k$ and $\epsilon_k$ represent the cross section and detection efficiency of the $k$-th background process in the Bkg-Exc category. To estimate the associated systematic uncertainty, each $\sigma_k$ is varied by its total uncertainty $\Delta\sigma_k$ (i.e., changed by $\pm1\sigma$), and the fit is repeated. The largest resulting variation in the extracted signal yield is taken as the systematic uncertainty due to the fixed Bkg-Exc background contribution.  
\end{itemize}
For the fitting method, the systematic uncertainties related to the fit procedure are treated as additive. The largest upper limit on the number of signal events, obtained from different fit conditions is taken. In addition, the multiplicative systematic uncertainty is incorporated by convolving the likelihood profile with respect to the signal yield corresponding to this most conservative result with a Gaussian function:
\begin{linenomath*}
\begin{equation*}
    L'(x) = \int_{0}^{1}L(x; N_{\rm sig}\bar{\epsilon}/\hat{\bar{\epsilon}})\exp[-\frac{(\bar{\epsilon}-\hat{\bar{\epsilon}})^2}{2\sigma_{\rm sys.}^2}]d\bar{\epsilon},
\end{equation*}
\end{linenomath*}
where $\hat{\bar{\epsilon}}$ is the nominal average detection efficiency, and $\sigma_{\rm sys.}$ is the total multiplicative systematic uncertainty~\cite{ref:Bayesianmethod,sigma}.

For the counting method, the additive systematic uncertainties include the $J/\psi$ signal region and the number of background events. The signal region is varied to $(3.075,~3.115)~{\rm GeV}/c^{2}$, $(3.085,~3.125)~{\rm GeV}/c^{2}$, the sideband region is changed to $(3.015,~3.055),~(3.135,~3.175)~{\rm GeV}/c^{2}$ or $(3.025,~3.065),~(3.145,~3.185)~{\rm GeV}/c^{2}$, and the number of Bkg-Exc events is changed with $\pm1\sigma$, and the largest upper limit of the number of signal events is chosen. The multiplicative systematic uncertainties are considered in the calculations of
the upper limit at $90\%$ C.L. by using the {\textsc{trolke}} package~\cite{t1} for the data samples taken at $\sqrt{s}~=~4467.06,~4527.14,\\4574.50,~{\rm and}~4611.86~{\rm MeV}$.

\section{Summary}

In summary, based on the data samples taken at $\sqrt{s}$ from $4395.4~{\rm MeV}$ to $4950.9~{\rm MeV}$, the $e^{+}e^{-}\to K^{0}K^{-}\pi^{+} J/\psi+c.c.$  process is observed for the first time.~Summing over all the data samples, the statistical significance is $9.4\sigma$.~For each of the 19 c.m.~energy points, the upper limit at the $90\%$ C.L. on the cross section for the process $\EE\to K^{0}K^{-}\pi^{+} J/\psi+c.c.$ is reported for the first time.~Additionally, no significant structure is observed in the cross section as a function of $\sqrt{s}$, and no obvious intermediate resonance states are detected among various combinations in the final state $K^{0}K^{-}\pi^+ J/\psi+c.c.$~The $Z_{cs}$ states are investigated in the $KJ/\psi$ and $K\pi J/\psi$ systems, yet no significant signals are observed within the existing sensitivity limits of the data samples.~More data samples are needed to investigate the vector $Y$, $Z_{c}$, and $Z_{cs}$ states.~This can be achieved with the upgraded BEPCII, which will plan to increase by a factor of three the luminosities at $XYZ$ states energies~\cite{Ablikim:2019hff}.

\acknowledgments

The BESIII Collaboration thanks the staff of BEPCII (https://cstr.cn/31109.02.BEPC) and the IHEP computing center for their strong support. This work is supported in part by the National Key R\&D Program of China under Contracts Nos. 2023YFA1606000 and 2023YFA1606704; the National Natural Science Foundation of China (NSFC) under Contracts Nos. 12375070, 11635010, 11935015, 11935016, 11935018, 12025502, 12035009, 12035013, 12061131003, 12192260, 12192261, 12192262, 12192263, 12192264, 12192265, 12221005, 12225509, 12235017, 12361141819; the Chinese Academy of Sciences (CAS) Large-Scale Scientific Facility Program; Joint Large Scale Scientific Facility Funds of the NSFC and CAS under Contracts Nos. U2032108; CAS under Contract No. YSBR-101; Shanghai Leading Talent Program of Eastern Talent Plan under Contract No. JLH5913002; the 100 Talents Program of CAS; the Institute of Nuclear and Particle Physics (INPAC) and the Shanghai Key Laboratory for Particle Physics and Cosmology; the European Research Council (ERC) under Contract No. 758462; the German Research Foundation (DFG) under Contract No. FOR5327; the Istituto Nazionale di Fisica Nucleare, Italy; the Knut and Alice Wallenberg Foundation under Contracts Nos. 2021.0174 and 2021.0299; the Ministry of Development of Turkey under Contract No. DPT2006K-120470; the National Research Foundation of Korea under Contract No. NRF-2022R1A2C1092335; the National Science and Technology Fund of Mongolia; the Polish National Science Centre under Contract No. 2024/53/B/ST2/00975; the Science and Technology Facilities Council (STFC, United Kingdom); the Swedish Research Council under Contract No. 2019.04595; and the U.S. Department of Energy under Contract No. DE-FG02-05ER41374.

\clearpage

\section*{The BESIII Collaboration}
\addcontentsline{toc}{section}{The BESIII Collaboration}
\begin{small}
%% Saved at => 2025-06-09
M.~Ablikim$^{1}$\BESIIIorcid{0000-0002-3935-619X},
M.~N.~Achasov$^{4,c}$\BESIIIorcid{0000-0002-9400-8622},
P.~Adlarson$^{77}$\BESIIIorcid{0000-0001-6280-3851},
X.~C.~Ai$^{82}$\BESIIIorcid{0000-0003-3856-2415},
R.~Aliberti$^{36}$\BESIIIorcid{0000-0003-3500-4012},
A.~Amoroso$^{76A,76C}$\BESIIIorcid{0000-0002-3095-8610},
Q.~An$^{59,73,a}$,
Y.~Bai$^{58}$\BESIIIorcid{0000-0001-6593-5665},
O.~Bakina$^{37}$\BESIIIorcid{0009-0005-0719-7461},
Y.~Ban$^{47,h}$\BESIIIorcid{0000-0002-1912-0374}
% ...
\\
\vspace{0.2cm}
(BESIII Collaboration)\\
\vspace{0.2cm} {\it
$^{1}$ Institute of High Energy Physics, Beijing 100049, People's Republic of China\\
% ...
$^{4}$ Budker Institute of Nuclear Physics SB RAS (BINP), Novosibirsk 630090, Russia\\
% ...
$^{36}$ Johannes Gutenberg University of Mainz, Johann-Joachim-Becher-Weg 45, D-55099 Mainz, Germany\\
$^{37}$ Joint Institute for Nuclear Research, 141980 Dubna, Moscow region, Russia\\
% ...
$^{47}$ Peking University, Beijing 100871, People's Republic of China\\
% ...
$^{58}$ Southeast University, Nanjing 211100, People's Republic of China\\
$^{59}$ State Key Laboratory of Particle Detection and Electronics, Beijing 100049, Hefei 230026, People's Republic of China\\
% ...
$^{73}$ University of Science and Technology of China, Hefei 230026, People's Republic of China\\
% ...
$^{76}$ University of Turin and INFN, (A)University of Turin, I-10125, Turin, Italy; (B)University of Eastern Piedmont, I-15121, Alessandria, Italy; (C)INFN, I-10125, Turin, Italy\\
$^{77}$ Uppsala University, Box 516, SE-75120 Uppsala, Sweden\\
% ...
$^{82}$ Zhengzhou University, Zhengzhou 450001, People's Republic of China\\

\vspace{0.2cm}
$^{a}$ Deceased\\
% ...
$^{c}$ Also at the Novosibirsk State University, Novosibirsk, 630090, Russia\\
% ...
$^{h}$ Also at State Key Laboratory of Nuclear Physics and Technology, Peking University, Beijing 100871, People's Republic of China\\
% ...

}
%% ends here %%
\vspace{0.4cm}
\end{small}

\end{document}